%% file: main.tex
\documentclass[a4paper,fleqn]{cas-sc}

\usepackage{placeins}
\usepackage[numbers]{natbib}

\def\tsc#1{\csdef{#1}{\textsc{\lowercase{#1}}\xspace}}
\tsc{WGM}
\tsc{QE}
\tsc{EP}
\tsc{PMS}
\tsc{BEC}
\tsc{DE}

\newcommand{\newIBM}{PC-IBM}
\newcommand{\oldIBM}{classical IBM}

\begin{document}
\let\WriteBookmarks\relax
\def\floatpagepagefraction{1}
\def\textpagefraction{.001}
\shorttitle{Physically consistent immersed boundary method}
\shortauthors{Hausmann, Elmestikawy, and van Wachem}
\renewcommand{\printorcid}{}
\title [mode = title]{Physically consistent immersed boundary method: a framework for predicting hydrodynamic forces on particles with coarse meshes}                      



\author[1]{Max Hausmann}
\author[1]{Hani Elmestikawy}
\address[1]{Chair of Mechanical Process Engineering, Otto-von-Guericke-Universität Magdeburg, Universitätsplatz 2, 39106 Magdeburg, Germany}
\author[1]{Berend {van Wachem}}

\begin{abstract}
In the present paper, a fluid-particle coupling method is directly derived from the Navier-Stokes equations (NSE) by applying the concept of volume-filtering, yielding a physically consistent methodology to incorporate solid wall boundary conditions in the volume-filtered flow solution, thereby allowing to solve the governing flow equations on non-body conforming meshes. The resulting methodology possesses similarities with a continuous forcing immersed boundary method (IBM) and is, therefore, termed physically consistent IBM (PC-IBM). Based on the recent findings of \citet{Hausmann2024a}, the closures arising in the volume-filtered NSE are closed by suitable models or even expressed analytically. The \newIBM{} is fully compatible with the large eddy simulation framework, as the volume-filtered NSE converge to the filtered NSE away from solid boundaries. The potential of the \newIBM{} is demonstrated by means of different particle-laden flow applications, including a dense packing of fixed particles and periodic settling of 500 particles. The new methodology turns out to be capable of accurately predicting the fluid forces on particles with relatively coarse mesh resolutions. For the flow around isolated spheres, a spatial resolution of six fluid mesh cells per diameter is sufficient to predict the drag force with less than 10\% deviation from highly resolved simulation for the investigated particle Reynolds numbers ranging from 10 to 250. 
\end{abstract}


\maketitle

\input{Sections/introduction}
\input{Sections/method}

\input{Sections/results}
\input{Sections/conclusions}
\input{Sections/acknowledgements}
\input{Sections/appendix}

\printcredits

\bibliographystyle{model1-num-names}



\end{document}

%% file: Sections/introduction.tex
\section{Introduction}
Although the available computing power is continuously growing, a detailed resolution of the flow features is still out of reach for the vast majority of applications. A significant portion of the computational resources is typically needed to resolve the complex fluid dynamics in the close vicinity of solid walls of almost any kind. A particularly complex boundary type of high practical relevance are solid particles, which possess curved surfaces and typically move in the considered frame of reference. The fluid dynamics are coupled with the particle dynamics via the boundary conditions at the particle surface. When the governing equations for the fluid motion are solved numerically on a mesh, a classical enforcement of boundary conditions requires degrees of freedom of the fluid mesh at the particle surface, i.e., a body fitted fluid mesh. Flow solvers employing body fitted fluid meshes are widely used \cite{Hirt1974,Benek1985,Thompson1974}. However, when the boundaries are moving, body fitted fluid meshes require frequent mesh adaption or remeshing, which is complicated, computationally expensive and may lead to low quality mesh cells reducing the accuracy. A common alternative is the immersed boundary method (IBM), which is initially introduced by \citet{Peskin1972,Peskin1977}. \\
According to \citet{Mittal2005}, the IBM approach involves a wide range of methodologies including cut-cell methods \cite{Clarke1985} and ghost-cell methods \cite{Majumdar2001}, which directly enforce the boundary conditions on the fluid mesh. The methodology proposed in the present paper, however, is closely related to the continuous forcing IBM \cite{Peskin1972}. In continuous forcing IBM, the fluid flow is solved in the whole computational domain, including the inside of the particle, and the boundary conditions at the particle surface are enforced by applying suitable momentum sources to the fluid momentum equation. The fluid momentum equation is evaluated at points at the particle surface, that are referred to as Lagrangian surface markers, where the fluid velocity is already known from the boundary conditions. A momentum source can be computed such that the desired velocity at the Lagrangian surface marker is achieved. However, after discretization of the momentum equation the fluid mesh cells do generally not coincide with the Lagrangian surface markers. Therefore, flow quantities have to be interpolated to the Lagrangian surface markers and the resulting momentum sources have to be spread to the Eulerian fluid mesh. Since the initial proposal of the continuous forcing IBM by \citet{Peskin1972}, several improvements have been made that enable relatively accurate simulations of particle-laden flows \cite{Uhlmann2005,Kempe2012,Roma1999,Pinelli2010,Breugem2012,Zhou2021}. Nevertheless, the interpolation and the spreading of variables governing the fluid flow introduce physical inconsistencies when the mesh resolution is finite. In practice, the continuous forcing IBM is only accurate with high spatial resolution \cite{AbdolAzis2018}. Furthermore, the continuous forcing IBM predicts a fluid motion inside the particle, which has no physical meaning. \\
In the present paper, we derive a fluid-particle coupling method directly from the Navier-Stokes equations (NSE) by volume-filtering that treats each term with a physically sound closure model. We refer to the newly proposed method as physically consistent IBM (PC-IBM). The \newIBM{} is based on volume-filtering of the NSE, which is originally introduced by \citet{Anderson1967}, and is commonly applied to particle-laden flows \cite{Capecelatro2013,Subramaniam2022}. Volume-filtering is closely related to the filtering applied in large eddy simulations (LES), but the convolution integrand is weighted with the fluid indicator function \cite{Anderson1967}. Applying the volume-filtering to the NSE gives rise to additional terms that require closures. In a recent study, \citet{Hausmann2024a} provide an analytical expression for the viscous closure, which arises because the operation of volume-filtering and the spatial derivative in the viscous term do not commute. Furthermore, \citet{Hausmann2024a}  also derive a model for the subfilter stress closure arising from volume-filtering the non-linear advective term in the NSE, which is shown to be accurate for small to moderate filter widths relative to the size of the particle. The third unclosed term in the volume-filtered NSE is a particle-momentum source represented by a particle surface integral of the fluid stresses weighted with the filter kernel. In the proposed \newIBM{}, this surface integral is discretized by means of Lagrangian surface markers, which is analogous to a continuous forcing IBM. An essential consequence of the volume-filtering is that the boundary conditions required for the volume-filtered velocity and pressure fields differ from the boundary conditions of the unfiltered flow field. The boundary conditions of the volume-filtered flow field can be obtained by explicitly filtering the boundary conditions of the unfiltered flow field, which results in a slip and transpiration boundary condition at the particle surface, i.e., a volume-filtered fluid velocity along and through the solid boundary. Slip and transpiration velocity boundary conditions at solid walls are by no means a new concept and are commonly applied in LES with wall modeling \cite{Bose2018,Bose2014,Bae2018}. \\
With the newly proposed \newIBM{}, it is shown that volume-filtering is a powerful tool that inherently links the concept of slip and transpiration velocity at the particle surface from wall modeled LES with the idea of enforcing velocity boundary conditions via momentum sources from continuous forcing IBM. Since the volume-filtering converges towards standard spatial filtering away from solid boundaries, the \newIBM{} is fully compatible with LES without further interventions. In fact, the model for the subfilter stress tensor employed in the \newIBM{}, such as the one proposed by \citet{Hausmann2024a}, converges to the so-called non-linear gradient model \citep{Liu1994b,Borue1998}, a well known single-phase flow subgrid-scale model, away from solid boundaries. The consistent treatment of filtering makes the \newIBM{} particularly suitable for the use of coarse meshes, analogous to LES.  \\
To the best of the authors' knowledge, the volume-filtering IBM proposed by \citet{Dave2023} is the only method that applies the ideas of volume-filtering to the framework of IBM. However, \citet{Dave2023} report large errors and instabilities originating from the computation of the advective term in the volume-filtered NSE, which they resolve by defining a second fluid phase inside the particle. Since the fluid phase inside the particle has the same density and viscosity, the resulting governing equations for the mixture are essentially similar to the single-phase LES equations and, their resulting numerical method is, strictly speaking, not based on volume-filtering. Moreover, the subfilter stress closure and the viscous closure are neglected in the method proposed by \citet{Dave2023}. It can be shown that the viscous closure is essential to preserve Galilean invariance of the volume-filtered NSE when the boundary is moving within the considered frame of reference \cite{Hausmann2024a}. Furthermore, \citet{Dave2023} apply no-slip boundary conditions at the particle surface, whereas it would be consistent to apply the boundary conditions for the (volume-)filtered fluid velocity (slip and transpiration), and the fact that the interpolation of Eulerian flow quantities to the Lagrangian surface markers constitutes another filtering operation is neglected. \\
The remainder of the paper is organized as follows. In section \ref{sec:methodology}, the concept of volume-filtering is introduced and the arising closures are briefly discussed. Furthermore, the discretization of the volume-filtered NSE is discussed and the details of the \newIBM{} are provided. In particular, we discuss the equivalence of interpolation and filtering operations, how the volume-filtered fluid momentum balance is evaluated at the particle surface, and how the slip and transpiration boundary conditions for the volume-filtered fluid velocity at the particle surface are obtained. The essential steps to convert a standard continuous forcing IBM to a \newIBM{} are described subsequently. In section \ref{sec:results}, the validation of the \newIBM{} is done based on simulations of the flow around an isolated sphere at different particle Reynolds numbers, the flow around arranged arrays and a dense random packing of spheres, an isolated settling sphere, and the settling of 500 spheres in a fully periodic domain. Section \ref{sec:conclusions} provides a summary and conclusions.

%% file: Sections/method.tex
\section{Methodology}
\label{sec:methodology}

\subsection{Volume-filtered equations}
\label{ssec:volumefilteredequations}
\begin{figure}
    \centering
    \includegraphics{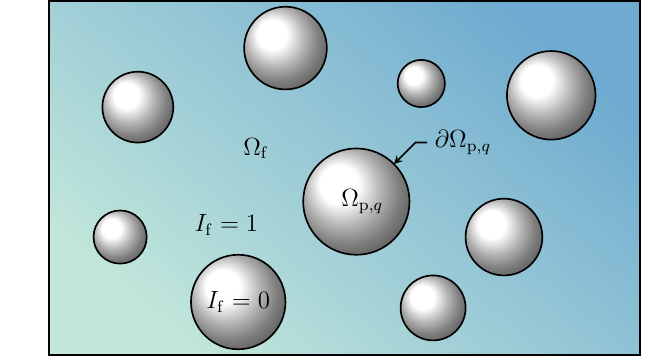}
    \caption{Sketch of a particle-laden flow as it is considered in the present study. The flow domain in blue is indicated with $\Omega_{\mathrm{f}}$ and the domain of the particle with the index $q$ with $\Omega_{\mathrm{p},q}$. The surface of the particle is referred to as $\partial \Omega_{\mathrm{p},q}$. The fluid indicator function, $I_{\mathrm{f}}$, is one in the fluid and zero inside the particles.}
    \label{fig:sketchdomain}
\end{figure}

We consider a fluid flow that contains one or more solid boundaries that potentially move, which we refer to as particles. Figure \ref{fig:sketchdomain} shows a sketch of a domain of a particle-laden flow. A particle with the index $q$ occupies the particle domain $\Omega_{\mathrm{p},q}$, whereas the fluid occupies $\Omega_{\mathrm{f}}$. The motion of the incompressible fluid of density $\rho_{\mathrm{f}}$, and constant dynamic viscosity, $\mu_{\mathrm{f}}$ is governed by the NSE
\begin{align}
   \label{eq:fluidcontinuity}
   \dfrac{\partial u_i}{\partial x_i} &= 0, \\
   \label{eq:fluidmomentum}
   \rho_{\mathrm{f}}\dfrac{\partial u_i}{\partial t} + \rho_{\mathrm{f}} \dfrac{\partial u_i u_j}{\partial x_j} &= - \dfrac{\partial p}{\partial x_i} + \mu_{\mathrm{f}} \dfrac{\partial}{\partial x_j}\left[\left(\dfrac{\partial u_i}{\partial x_j}+\dfrac{\partial u_j}{\partial x_i}\right)\right],
\end{align}
where $p$ is the pressure field and $u_i$ is the fluid velocity field. At the surface of the particle with the index $q$, $\partial \Omega_{\mathrm{p},q}$, the fluid velocity vector equals the velocity vector at the particle boundary consisting of a translational particle velocity, $v_{q,i}$, and particle angular velocity, $\omega_{q,i}$, such that
\begin{align}
\label{eq:boundarycondition}
   \boldsymbol{u}(\boldsymbol{x}_{\partial \Omega_{\mathrm{p},q}}) = \boldsymbol{v}_{q} + \boldsymbol{\omega}_q \times (\boldsymbol{x}_{\partial \Omega_{\mathrm{p},q}}-\boldsymbol{x}_{\mathrm{p},q}),
\end{align}
where $\boldsymbol{x}_{\partial \Omega_{\mathrm{p},q}}$ is the position of a point at the particle surface and $\boldsymbol{x}_{\mathrm{p},q}$ is the center of mass of the particle. \\
Resolving the geometry of the particle and the potentially large spatial gradients in velocity and pressure, especially but not exclusively near the particle surface, requires a high spatial resolution and, therefore, leads to high computational costs. Although the small flow structures are responsible for essential processes in the flow, such as kinetic energy dissipation, kinetic energy production, and nonlocal vortex dynamics \cite{Pope2000,Madylam2009}, the small flow structures are of less interest from a macroscopic point of view. In LES, accounting for the effect of the small flow structures by appropriate closure modeling, justifies the solution of only the large scales, i.e., the spatially filtered NSE (see, e.g., \citet{Sagaut2005}). \\
The generalization of spatial filtering of multiple phases applied to a flow quantity $\varPhi$, also referred to as volume-filtering, as originally introduced by \citet{Anderson1967}, and is defined as the following convolution integral over the volume of the domain
\begin{align}
\label{eq:volumefiltering}
    \epsilon_{\mathrm{f}}(\boldsymbol{x}) \overline{\varPhi}(\boldsymbol{x}) = \int\displaylimits_{\Omega} I_{\mathrm{f}}(\boldsymbol{y}) \varPhi(\boldsymbol{y})g(|\boldsymbol{x}-\boldsymbol{y}|) \mathrm{d}V_y , 
\end{align}
where $\Omega=\Omega_{\mathrm{f}}\cup \Omega_{\mathrm{p}}$ is the union of the domains occupied by the fluid and the particle, and $\epsilon_{\mathrm{f}}$ is the fluid volume fraction
\begin{align}
    \epsilon_{\mathrm{f}}(\boldsymbol{x})  = \int\displaylimits_{\Omega} I_{\mathrm{f}}(\boldsymbol{y})g(|\boldsymbol{x}-\boldsymbol{y}|) \mathrm{d}V_y.
\end{align}
Note that time dependencies are omitted for conciseness. The uniform and symmetric filter kernel, $g$, has to satisfy
\begin{align}
    \int \displaylimits_{\Omega} g(\boldsymbol{|x|})\mathrm{d}V_x = 1,
\end{align}
and the fluid indicator function, $I_\mathrm{f}$, is defined as
\begin{align}
    I_\mathrm{f}(\boldsymbol{x})=\begin{cases}
        1 & \text{if } \boldsymbol{x} \in \Omega_\mathrm{f}\\
        0 & \text{else }.
    \end{cases}
\end{align}
Quantities with a bar are referred to as volume-filtered. Although the filtering is purely spatial, the flow field is implicitly filtered in time because small flow scales, that are filtered out, change with a higher frequency than large flow scales \cite{Sagaut2005}. Since the kernel that is associated with the implicit temporal filtering depends on the frequency of the filtered out small scales, the exact influence of the temporal filtering is unknown and typically ignored when spatial filtering is applied, such as in LES \cite{Sagaut2005}. \\
Without further assumptions, the volume-filtering of the NSE gives \cite{Hausmann2024a}
\begin{align}
\label{eq:unclosedcontinuity}
    \dfrac{\partial \epsilon_{\mathrm{f}}}{ \partial t} + \dfrac{\partial}{\partial x_i}(\epsilon_{\mathrm{f}}\Bar{u}_i) &= 0, \\
    \label{eq:unclosedmomentum}
    \rho_{\mathrm{f}}\dfrac{\partial \epsilon_{\mathrm{f}}\Bar{u}_i}{\partial t} + \rho_{\mathrm{f}}\dfrac{\partial}{\partial x_j}(\epsilon_{\mathrm{f}}\overline{u_i u_j}) &= -\dfrac{\partial \epsilon_{\mathrm{f}}\Bar{p}}{\partial x_i}+\mu_{\mathrm{f}}\dfrac{\partial}{\partial x_j}\left[ \epsilon_{\mathrm{f}}\left( \overline{\dfrac{\partial u_i}{\partial x_j}} + \overline{\dfrac{\partial u_j}{\partial x_i}} \right) \right] - s_i,
\end{align}
where the overbar denotes the filtered variables, as defined by equation \eqref{eq:volumefiltering}, and the particle momentum source, $s_i$, is given as the following integral over the particle surfaces
\begin{align}
\label{eq:particlemomentumsource}
    s_i = \int\displaylimits_{\partial\Omega_{\mathrm{p}}}g(|\boldsymbol{x}-\boldsymbol{y}|)\left(-p \delta_{ij} + \mu_{\mathrm{f}} \left(\dfrac{\partial u_i}{\partial y_j}+\dfrac{\partial u_j}{\partial y_i}\right)\right)n_j\mathrm{d}A_y.
\end{align}
The normal vector, $n_j$, points outwards from the surface of the particle under consideration. For conciseness of following derivations in this paper, we define the fluid stress vector, $\Sigma_i$, as the contraction of the fluid stress tensor with the normal vector
\begin{align}
    \Sigma_i = \left(-p \delta_{ij} + \mu_{\mathrm{f}} \left(\dfrac{\partial u_i}{\partial y_j}+\dfrac{\partial u_j}{\partial y_i}\right)\right)n_j.
\end{align}
The volume-filtered NSE \eqref{eq:unclosedcontinuity} and \eqref{eq:unclosedmomentum} are not closed, as they contain the unfiltered pressure and the unfiltered fluid velocity field. According to \citet{Hausmann2024a}, the volume-filtered momentum equation can be written without loss of generality as follows
\begin{align}
    \rho_{\mathrm{f}}\dfrac{\partial \epsilon_{\mathrm{f}}\Bar{u}_i}{\partial t} + \rho_{\mathrm{f}}\dfrac{\partial}{\partial x_j}(\epsilon_{\mathrm{f}}\Bar{u}_i\epsilon_{\mathrm{f}}\Bar{u}_j) &= -\dfrac{\partial \epsilon_{\mathrm{f}}\Bar{p}}{\partial x_i}+\mu_{\mathrm{f}} \dfrac{\partial^2\epsilon_{\mathrm{f}}\Bar{u}_i}{\partial x_j \partial x_j} -s_i\color{black} + \mu_{\mathrm{f}} \mathcal{E}_i- \rho_{\mathrm{f}}\dfrac{\partial }{\partial x_j}\tau_{\mathrm{sfs},ij},
\end{align}
where the viscous closure, $\mathcal{E}_i$, originates from the modification of the viscous term, and the subfilter stress tensor $\tau_{\mathrm{sfs},ij}$ originates from modifying the advective term. The viscous closure is analytically derived in \citet{Hausmann2024a} for spherical particles to be
\begin{align}
\label{eq:viscousclosure}
    \mathcal{E}_i=\sum\displaylimits_q v_{q,i}\dfrac{\partial^2 \epsilon_{\mathrm{p},q}}{\partial x_j \partial x_j},
\end{align}
where $v_{q,i}$ is the velocity vector of the particle with index $q$. The particle volume fraction of particle $q$ is defined as 
\begin{align}
    \epsilon_{\mathrm{p},q}(\boldsymbol{x})  = \int\displaylimits_{\Omega} I_{\mathrm{p},q}(\boldsymbol{y})g(|\boldsymbol{x}-\boldsymbol{y}|) \mathrm{d}V_y,
\end{align}
where $I_{\mathrm{p},q}$ is the indicator function of particle $q$, which is one inside particle $q$ and zero elsewhere. The subfilter stress closure requires modeling since no analytical expression exists. While a large variety of models for the subfilter stress tensor exist in single-phase flows, to the best of the authors' knowledge the only existing model in the context of volume-filtering of incompressible flows is the model from \citet{Hausmann2024a}. According to this model, the subfilter stress tensor contains a Galilean invariance contribution, $\tau_{\mathrm{sfs},ij}^{\mathrm{G}}$, that cancels out all additional terms of the momentum equations that arise when the frame of reference is changed, but does not model the interscale momentum and energy transfer. To cancel out the terms that arise from a change of the frame of reference, the minimal expression for $\tau_{\mathrm{sfs},ij}^{\mathrm{G}}$ is
\begin{align}
    \tau_{\mathrm{sfs},ij}^{\mathrm{G}} = \sum_{q}\left[ \epsilon_{\mathrm{p},q} v_{q,j}\epsilon_{\mathrm{f}} \Bar{u}_i + \epsilon_{\mathrm{p},q} v_{q,i}\epsilon_{\mathrm{f}} \Bar{u}_j - \epsilon_{\mathrm{p},q} v_{q,i} \epsilon_{\mathrm{f}} v_{q,j}\right].
\end{align}
In addition to the Galilean invariance contribution, there is a contribution that models the non-linear subfilter interaction, such that the modeled subfilter stress tensor $\tau_{\mathrm{sfs},ij}^{\mathrm{mod}}$ is given as
\begin{align}
\label{eq:subfilterstressclosure}
    \tau_{\mathrm{sfs},ij}^{\mathrm{mod}} = \tau_{\mathrm{sfs},ij}^{\mathrm{G}} + \sigma^2 \left[ \dfrac{\partial \epsilon_{\mathrm{f}} \Bar{u}_i }{\partial x_k} + \sum_q v_{q,i} \dfrac{\partial \epsilon_{\mathrm{p},q}}{\partial x_k}  \right]\left[ \dfrac{\partial \epsilon_{\mathrm{f}} \Bar{u}_j }{\partial x_k} + \sum_q v_{q,j}\dfrac{\partial \epsilon_{\mathrm{p},q}}{\partial x_k}  \right],
\end{align}
which is derived for the Gaussian filter kernel
\begin{align}
    g(\boldsymbol{x}) = \dfrac{1}{(2\pi \sigma^2)^{3/2}}\exp\left( -\dfrac{|\boldsymbol{x}|^2}{2 \sigma^2} \right).
\end{align}
For fixed particles, i.e., $v_{q,i}=0$, the viscous closure and the subfilter stress tensor significantly simplify to
\begin{align}
    \mathcal{E}_i=0, \phantom{0} \tau_{\mathrm{sfs},ij}^{\mathrm{mod}} = \sigma^2\dfrac{\partial \epsilon_{\mathrm{f}} \Bar{u}_i }{\partial x_k}\dfrac{\partial \epsilon_{\mathrm{f}} \Bar{u}_j }{\partial x_k}.
\end{align}
The treatment of the remaining unclosed term, the particle momentum source, $s_i$, links the presented methodology to the continuous forcing IBM and is discussed in the subsequent sections of this paper. \\
It is noteworthy that the volume-filtered NSE can be transformed such that they are similar to the unfiltered NSE with additional source terms. Defining the transformed variables $u_{\epsilon,i}=\epsilon_{\mathrm{f}}\Bar{u}_i$ and  $p_{\epsilon}=\epsilon_{\mathrm{f}}\Bar{p}$, the volume-filtered NSE reduce to
\begin{align}
\label{eq:reducedNSEconstinuity}
    \dfrac{\partial \epsilon_{\mathrm{f}}}{ \partial t} + \dfrac{\partial u_{\epsilon,i}}{\partial x_i} &= 0, \\
\label{eq:reducedNSEmomentum}
    \rho_{\mathrm{f}}\dfrac{\partial u_{\epsilon,i}}{\partial t} + \rho_{\mathrm{f}}\dfrac{\partial}{\partial x_j}(u_{\epsilon,i}u_{\epsilon,j}) &= -\dfrac{\partial p_{\epsilon}}{\partial x_i} +\mu_{\mathrm{f}} \dfrac{\partial^2u_{\epsilon,i}}{\partial x_j \partial x_j} -s_i\color{black} + \mu_{\mathrm{f}} \mathcal{E}_i- \rho_{\mathrm{f}}\dfrac{\partial }{\partial x_j}\tau_{\mathrm{sfs},ij},
\end{align}
which can be solved with any standard incompressible flow solver, but the explicitly known mass source, $\partial\epsilon_{\mathrm{f}} / \partial t$, and momentum source terms have to be added. In the case of fixed particles, the mass source vanishes. Note that the solution of the system consisting of equations \eqref{eq:reducedNSEconstinuity}-\eqref{eq:reducedNSEmomentum} can be solved with a projection method analog to the solution of the incompressible NSE with variable density as described, for instance, in \citet{Almgren1998}.

\subsection{Discretization of the governing equations}
In this work, the governing fluid equations \eqref{eq:reducedNSEconstinuity}-\eqref{eq:reducedNSEmomentum} are discretized using a finite volume method on a fixed Eulerian fluid mesh with second order spatial and temporal accuracy. The flow solver employs a collocated variable arrangement and a monolithic solution of the continuity and momentum equation using momentum weighted interpolation. An advecting velocity is defined as the sum of the interpolated cell centered velocity and a pressure correction term, which is applied in the continuity equation and the advection term of the momentum equation. The transient term in the momentum equation is discretized with a second order backward Euler (BDF2) scheme, in the advective term we use a SuperBee flux limiter with a slope $L=2$ of the flux limiter with respect to the gradient ratio (see \citet{Denner2015a} for details), and the pressure and diffusive term are discretized with central differences, respectively. Details on the momentum weighted interpolation and the discretization of all terms can be found in \citet{Denner2014a,Bartholomew2018}, and \citet{Denner2020}. In the present study, the flow is solved on a Cartesian mesh, although the methodology can be easily extended to a solver using any type of mesh. \\
In the newly proposed \newIBM{}, it has to be distinguished between different filter widths. The filter width of the Eulerian quantities, such as the fluid velocity and pressure field, is referred to as $\sigma_{\mathrm{E}}$. The corresponding Gaussian filter kernel possessing a standard deviation of $\sigma_{\mathrm{E}}$ is indicated as $g_{\mathrm{E}}$. Throughout the paper, the filter width refers to the standard deviation of the Gaussian filter kernel.  \\
\begin{figure}
    \centering
    \hspace{2cm}
    \includegraphics[scale=0.7]{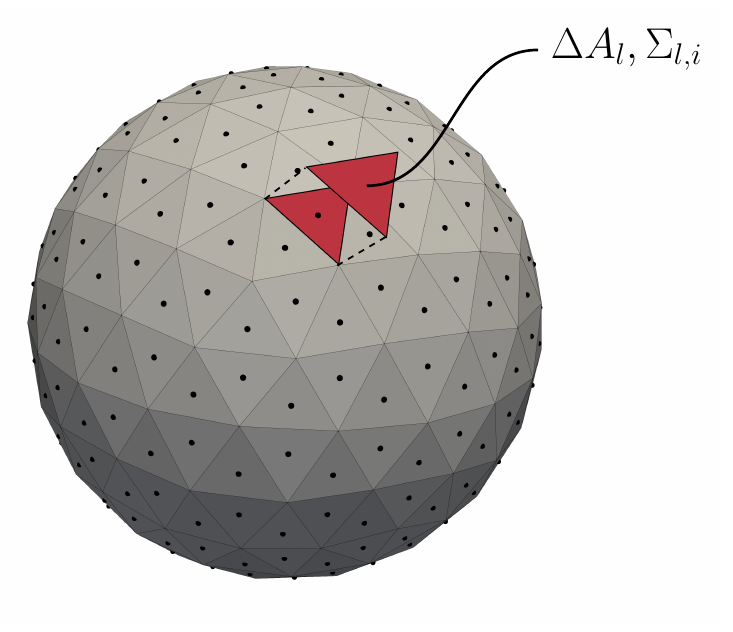}
    \caption{Visualization of the discretization of the particle surface. The points on the surface indicate the Lagrangian surface markers, which are the centers of the surrounding triangles. Each surface triangle possesses an associated surface area, $\Delta A_l$, and a stress vector, $\Sigma_{l,i}$. }
    \label{fig:sketchtriangulatedsurface}
\end{figure}
Equation \eqref{eq:reducedNSEmomentum} contains a term that remains to be discretized, the particle momentum source, $s_i$. The surface integral in equation \eqref{eq:particlemomentumsource} is discretized with a finite sum of $N_{\mathrm{s}}$ triangles at the particle surface. The discretization of the particle surface is visualized in figure \ref{fig:sketchtriangulatedsurface}, where the Lagrangian surface markers, i.e., the centers of the surface triangles, and the corresponding surface triangles are highlighted. With this, the discretized particle momentum source is
\begin{align}
 \label{eq:particlemomentumsourcediscrete}
     s_i(\boldsymbol{x}) \approx \sum_{l=1}^{N_{\mathrm{s}}} g_{\mathrm{E}}(|\boldsymbol{x}-\boldsymbol{X}_l|)\Sigma_{l,i} \Delta A_l,
 \end{align}
where  $\boldsymbol{X}_l$ is the center of the surface triangle with the index $l$, $\Sigma_{l,i}$ is the fluid stress vector at the surface triangle, and $\Delta A_l$ is the area of the surface triangle with the index $l$.  Note that the discretization of the particle surface must not necessarily be uniform or done with triangles. However, it is important to be able to associate a surface area to each Lagrangian surface marker. \\
If the fluid stress vector at the surface triangles, $\Sigma_{l,i}$, can be determined, then the equation system consisting of equation \eqref{eq:reducedNSEconstinuity} and \eqref{eq:reducedNSEmomentum} is closed and can be solved numerically. 

\subsection{Physically consistent immersed boundary method}
We refer to an IBM as \oldIBM{} if the fluid velocity is interpreted as not being filtered (such as in \citet{Peskin1972,Peskin1977,Uhlmann2005,Kempe2012,AbdolAzis2018}), or it is interpreted as filtered without performing all necessary steps to obtain the complete volume-filtered equations \cite{Dave2023}. The important steps to obtain the correct governing equations for the filtered velocity field include enforcing a slip and transpiration fluid velocity at the particle surface, considering the closures of the volume-filtered momentum equation, and accounting for the equivalence of interpolation and filtering. It is shown in the remainder of the present section that a slip velocity at the particle surface and the equivalence of interpolation and filtering are consequences of interpreting the flow field as volume-filtered. \\
A \oldIBM{} is, at most, consistent in a numerical sense, since it converges to the solution of the flow over a particle as the grid is refined. In the converged limit, the feedback force from the Lagrangian surface markers on the particle surface to the flow is spread with a Dirac delta distribution. The spreading of the force from the Lagrangian surface marker on the particle onto the Eulerian fluid mesh using a kernel of finite size, however, lacks a physical justification if the velocity field is not interpreted as filtered. Analogously, solving for a turbulent flow with a too coarse discretization in time and space, instead of resolving all turbulent length- and time scales with a direct numerical simulation (DNS), is physically unjustified. In both cases the interactions with the small scales, which are not resolved, are not accounted for. However, interpreting the turbulent flow with the coarse discretization as a filtered flow field and accounting for the effect of the subgrid-scale flow features by appropriate closures in the solved equations is known as LES and is physically consistent. Analog to the relation between LES and DNS, we propose the \newIBM{} as a physically consistent alternative to \oldIBM{} if a full resolution of all geometric and flow features is out of reach. The \newIBM{} converges towards the \oldIBM{} as the numerical grid is refined, similar to a LES converging towards a DNS under grid refinement. \\
The \newIBM{} solves for the fluid stresses at the surface triangles that lead to the desired volume-filtered fluid velocities at the Lagrangian surface markers in the center of the surface triangles. 

\subsubsection{Interpolation is equivalent to filtering}
\label{ssec:interpolationequalsspreading}
\begin{figure}
    \centering
    \hspace{-1.5cm}
    \includegraphics[scale=0.9]{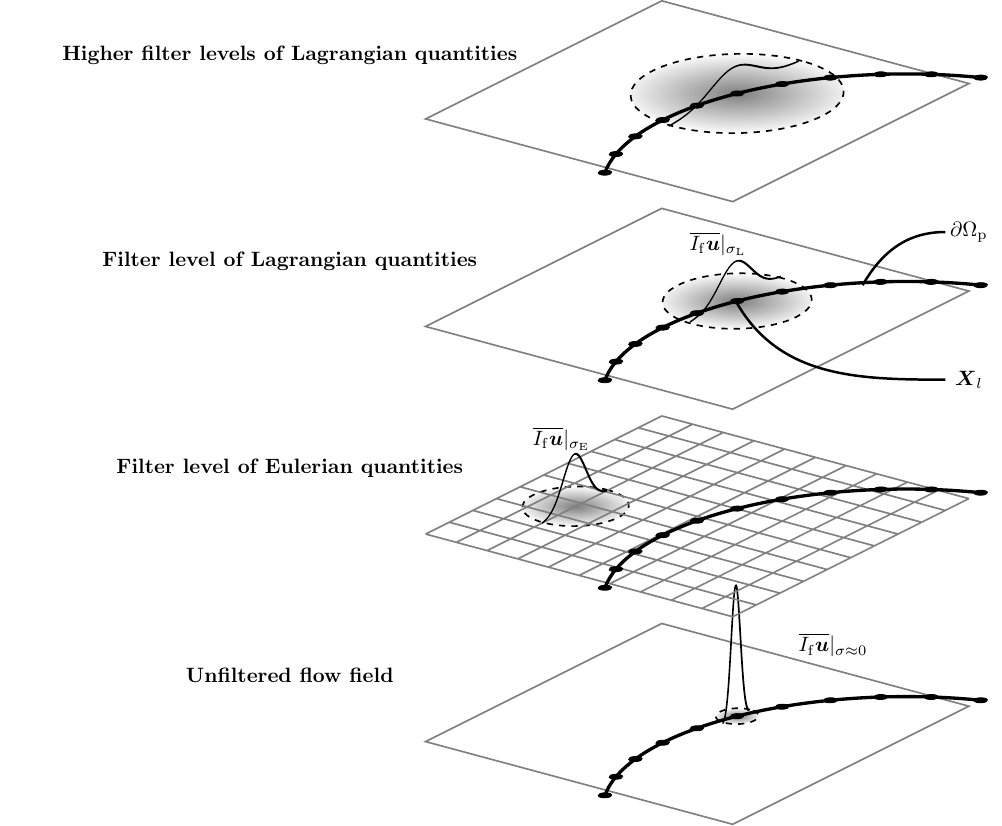}
    \caption{Two-dimensional visualization of the filter levels that are considered in the \newIBM{}. The particle surface, $\partial \Omega_{\mathrm{p}}$, and one Lagrangian surface marker with the position $\boldsymbol{X}_l$ are shown. Gaussian functions are drawn that indicate the filter kernel, $g$, of the corresponding filtered quantity. The lowest level represents the unfiltered solution, which is unknown in the \newIBM{}. The second level represents Eulerian quantities in the \newIBM{} with a corresponding filter width $\sigma_{\mathrm{E}}$ and the third level Lagrangian quantities in the \newIBM{} with a larger filter width $\sigma_{\mathrm{L}}$ originating from the interpolation to the particle surface. The highest level indicates any larger filter width of Lagrangian quantities that can be easily obtained by explicit filtering or interpolation of Eulerian quantities. }
    \label{fig:sketchfiltering}
\end{figure}

We introduce a notation to indicate the filter width of a volume-filtered quantity. A flow quantity, $\varPhi$, weighted with the fluid indicator function, $I_{\mathrm{f}}$, and volume-filtered with a Gaussian filter kernel with a standard deviation, $\sigma$, is defined as 
\begin{align}
\label{eq:volumefilteringprecise}
    \overline{I_{\mathrm{f}}\varPhi}\vert_{\sigma}(\boldsymbol{x}) = \int\displaylimits_{\Omega} I_{\mathrm{f}}(\boldsymbol{y}) \varPhi(\boldsymbol{y})g(|\boldsymbol{x}-\boldsymbol{y}|) \mathrm{d}V_y.
\end{align}
Note that in the case of a Gaussian filter kernel, filtering twice with different filter widths, $\sigma_{\mathrm{a}}$ and $\sigma_{\mathrm{b}}$, can be expressed as a single filtering operation with equivalent filter width $\sqrt{\sigma_{\mathrm{a}}^2+\sigma_{\mathrm{b}}^2}$ \cite{Johnson2021,Hausmann2024a},
\begin{align}
\label{eq:subsequentfiltering}
\overline{\overline{I_{\mathrm{f}}u_i}\vert_{\sigma_{\mathrm{a}}}}\vert_{\sigma_{b}} = \overline{I_{\mathrm{f}}u_i}\vert_{\sqrt{\sigma_{\mathrm{a}}^2+\sigma_{\mathrm{b}}^2}}.
\end{align}
The Lagrangian surface markers in the center of the surface triangles generally do not coincide with the centers of the Eulerian fluid mesh cells where the flow quantities are known. Therefore, flow quantities have to be interpolated from the Eulerian fluid mesh cells to the Lagrangian surface markers. The interpolation of quantities from the cell centers of the Eulerian fluid mesh, $\boldsymbol{y}_m$, to the Lagrangian surface markers, $\boldsymbol{X}_l$, with the Gaussian interpolation kernel $g_{\delta}$ of standard deviation $\sigma_{\delta}$ is a discrete approximation of the following integral
\begin{align}
\label{eq:interpolation}
    \int g_{\delta}(|\boldsymbol{X}_l - \boldsymbol{y}|)\overline{I_{\mathrm{f}}u_i}\vert_{\sigma_{\mathrm{E}}}(\boldsymbol{y})\mathrm{d}V_y \approx \sum_m \overline{I_{\mathrm{f}}u_i}\vert_{\sigma_{\mathrm{E}}}(\boldsymbol{y}_m) g_{\delta}(|\boldsymbol{X}_l - \boldsymbol{y}_m|) \Delta V_m,
\end{align}
where $\Delta V_m$ is the volume of an Eulerian fluid mesh cell and the index $m$ refers to all fluid mesh cells in which $g_{\delta}(|\boldsymbol{X}_l - \boldsymbol{y}_m|)$ significantly deviates from zero. Since the integral is equal to a filtering operation, 
\begin{align}
     \int g_{\delta}(|\boldsymbol{X}_l - \boldsymbol{y}|)\overline{I_{\mathrm{f}}u_i}\vert_{\sigma_{\mathrm{E}}}(\boldsymbol{y})\mathrm{d}V_y = \overline{\overline{I_{\mathrm{f}}u_i}\vert_{\sigma_{\mathrm{E}}}}\vert_{\sigma_{\delta}}(\boldsymbol{X}_l) = \overline{I_{\mathrm{f}}u_i}\vert_{\sqrt{\sigma_{\mathrm{E}}^2+\sigma_{\delta}^2}}(\boldsymbol{X}_l),
\end{align}
interpolation with a Gaussian kernel increases the filter width. Therefore, all flow quantities interpolated from the Eulerian fluid mesh to the Lagrangian surface markers exist at a filter width $\sigma_{\mathrm{L}} = \sqrt{\sigma_{\mathrm{E}}^2+\sigma_{\delta}^2}$. If the filter width for the interpolation is $\sigma_{\delta}=\sigma_{\mathrm{E}}$, the spreading weights of the force using equation \eqref{eq:particlemomentumsourcediscrete} are equal to the interpolation weights because the Gaussian kernels appearing in equation \eqref{eq:particlemomentumsourcediscrete} and equation \eqref{eq:interpolation} are the same ($g_{\delta}=g_{\mathrm{E}}$). \\
The different filter levels of the \newIBM{} are visualized in figure \ref{fig:sketchfiltering} and can be described from the lowest to the highest level as follows:
\begin{itemize}
    \item The lowest level represents the essentially unfiltered flow field, which corresponds to $\sigma_{\mathrm{E}}\rightarrow 0$ and $\sigma_{\delta} \rightarrow 0 $ and is unknown in the \newIBM{}. A very fine numerical fluid mesh would be required to approximate the unfiltered flow field. 
    \item The second level represents the volume-filtered solution of the \newIBM{} on the Eulerian fluid mesh with a filter width $\sigma_{\mathrm{E}}$. The filter width is chosen, such that the volume-filtered flow quantities can be resolved by the fluid mesh and it is imposed by the width of the Gaussian spreading kernel in the particle momentum source given in equation \eqref{eq:particlemomentumsourcediscrete}, the filter width applied in the model for the subfilter stress tensor equation \eqref{eq:subfilterstressclosure}, and the filter width applied in the viscous closure equation \eqref{eq:viscousclosure}.
    \item The third level represents Eulerian flow quantities that are interpolated to the Lagrangian surface markers, $\boldsymbol{X}_l$, with a Gaussian interpolation kernel, $g_{\delta}$, and, therefore, possess a larger filter width $\sigma_{\mathrm{L}}$.
    \item The \newIBM{} requires larger filter levels for Lagrangian quantities, which is represented by the highest level in figure \ref{fig:sketchfiltering} (details provided later in section \ref{ssec:desiredvelocity}). Larger filter levels of Lagrangian flow quantities can be easily obtained by increasing the standard deviation of the Gaussian interpolation kernel.
\end{itemize}

\subsubsection{Fluid momentum equation at the particle surface}
The fluid stress vector at the surface triangles is evaluated by considering the fluid momentum equation at the particle surface. The momentum equation \eqref{eq:reducedNSEmomentum} evaluated at the particle surface possesses two unknowns, the fluid stress vector, $\Sigma_{l,i}$, and the volume-filtered fluid velocity vector at the new time level. However, the volume-filtered fluid velocity of the next time step at the particle surface depends on the velocity of the interface at the triangle center, $u_{\mathrm{Int},l,i}$, where the index $l$ refers to the Lagrangian surface marker. Assuming a no-slip and no-penetration boundary condition for the unfiltered flow field at the particle surface, the unfiltered fluid velocity is given as 
\begin{align}
    u_i(\boldsymbol{X}_l) = u_{\mathrm{Int},l,i} = v_{q,i} + \omega_{q,i} \times (X_{l,i}-x_{\mathrm{p},q,i}).
\end{align}
Since Lagrangian quantities exist at a filter width $\sigma_{\mathrm{L}}>0$, the imposed fluid velocity is a desired volume-filtered fluid velocity, $(\overline{I_{\mathrm{f}}u_i}\vert_{\sigma_{\mathrm{L}}})_{\mathrm{des}}$, which is described in section \eqref{ssec:desiredvelocity}. The desired volume-filtered fluid velocity represents the slip along and the transpiration through the particle. A momentum source can be applied in a region close to the particle surface such that the desired volume-filtered fluid velocity is obtained in the next time step, which is conceptually similar to how the momentum source term is obtained in a \oldIBM{}. Interpolating the volume-filtered fluid momentum equation \eqref{eq:reducedNSEmomentum} with the Gaussian filter kernel, $g_{\delta}$, gives the fluid momentum balance at the Lagrangian surface marker $l$
\begin{align}  
\label{eq:momentumatsurface}
\rho_{\mathrm{f}}\dfrac{(\overline{I_{\mathrm{f}}u_i}\vert_{\sigma_{\mathrm{L}}})_{\mathrm{des}}(\boldsymbol{X}_l) - (\overline{I_{\mathrm{f}}u_i}\vert_{\sigma_{\mathrm{L}}})^n(\boldsymbol{X}_l)}{\Delta t} +\mathcal{A}_i^n(\boldsymbol{X}_l) = - \mathcal{P}_i^n(\boldsymbol{X}_l) + \mathcal{V}_i^n(\boldsymbol{X}_l) - \sum_{m=1}^{N_{\mathrm{s}}} g_{\mathrm{L}}(|\boldsymbol{X}_l-\boldsymbol{X}_m|)\Sigma_{m,i} \Delta A_m,
\end{align}
where $\mathcal{A}_i$ is the interpolated advective term including the subfilter stress closure, $\mathcal{P}_i$ is the interpolated pressure term, and $\mathcal{V}_i$ is the interpolated viscous term including the viscous closure. The last term on the right-hand side of equation \eqref{eq:momentumatsurface} contains a Gaussian kernel, $g_{\mathrm{L}}$, with a standard deviation $\sigma_{\mathrm{L}}$ and the superscript $n$ refers to quantities of the previous time step.\\
Equation \eqref{eq:momentumatsurface} is solved for every Lagrangian surface marker and, because of the last term on the right-hand, the solution for each Lagrangian surface marker depends on the solution of other Lagrangian surface markers within the support of $g_{\mathrm{L}}$. Therefore, equation \eqref{eq:momentumatsurface} can be reformulated into the equation system
\begin{align}
\label{eq:equationsystemsigma}
    A_{lm}\boldsymbol{\Sigma}_m=\boldsymbol{b}_l,
\end{align}
where the matrix $A_{lm}$ originates from the last term on the right-hand side of equation \eqref{eq:momentumatsurface} and contains the coefficients consisting of the product of the Gaussian, $g_{\mathrm{L}}$, and the surface areas of the triangles of the discretized particle surface, $\Delta A_m$. The vector $\boldsymbol{b}_l$ contains the remaining explicitly known terms of equation \eqref{eq:momentumatsurface}. The solution of the equation system \eqref{eq:equationsystemsigma} is not practicable as it essentially poses a defiltering operation without regularization. In other words, an inverse filtering operation is applied to a quantity that is not exactly the filtered quantity, but an approximation of it. As described in detail in \citet{Eyink2006}, such an inversion of a filtering operation is exact when the support of the Fourier transform of the filter kernel is infinite and when the filtered field is obtained by actual explicit filtering. The latter is not satisfied in the present case because $\boldsymbol{b}_l$ contains modeling and discretization errors. If the filtered field is an approximation, the unfiltered solution is generally unbounded. In practice, even small inaccuracies of the vector $\boldsymbol{b}_l$ of the equation system \eqref{eq:equationsystemsigma} amplify when the system is solved and no physically reasonable solution is obtained. \\
Another drawback is the high computational cost of inverting a system of the size of the number of Lagrangian surface markers, $N_{\mathrm{s}}$, especially for many particles. Therefore, we assume a fluid stress vector $\tilde{\Sigma}_{l,i}$ that is constant within the support of the Gaussian kernel $g_{\mathrm{L}}$ (since the support is infinite, we assume $\tilde{\Sigma}_{l,i}$ constant where the Gaussian significantly deviates from zero), which allows
\begin{align}
\label{eq:stressconstantwithingsupport}
    \sum_{m=1}^{N_{\mathrm{s}}} g_{\mathrm{L}}(|\boldsymbol{X}_l-\boldsymbol{X}_m|)\Sigma_{m,i} \Delta A_m = \tilde{\Sigma}_{l,i}\sum_{m=1}^{N_{\mathrm{s}}} g_{\mathrm{L}}(|\boldsymbol{X}_l-\boldsymbol{X}_m|) \Delta A_m.
\end{align}
This assumption allows to solve equation \eqref{eq:momentumatsurface} explicitly for every $\tilde{\Sigma}_{l,i}$. However, an error is introduced during the spreading of the particle momentum source onto the Eulerian fluid mesh, i.e., equation \eqref{eq:particlemomentumsourcediscrete}, since the following relation is only approximate
\begin{align}
    \sum_{l=1}^{N_{\mathrm{s}}} g_{\mathrm{E}}(|\boldsymbol{x}-\boldsymbol{X}_l|)\Sigma_{l,i} \Delta A_l \approx \sum_{l=1}^{N_{\mathrm{s}}} g_{\mathrm{E}}(|\boldsymbol{x}-\boldsymbol{X}_l|)\tilde{\Sigma}_{l,i} \Delta A_l.
\end{align}
Consequently, the obtained volume-filtered fluid velocity field deviates from the desired field. We refer to this fluid velocity deviation at the particle surface as the slip error. The assumption of equation \eqref{eq:stressconstantwithingsupport} is justified as long as the maximum deviation of the interpolated fluid velocity from the desired volume-filtered fluid velocity, i.e., the maximum slip error, is reasonably small. The smaller the filter width or the smoother $\Sigma_{l,i}$ varies across the particle surface, the more valid the replacement of $\Sigma_{l,i}$ with $\tilde{\Sigma}_{l,i}$. It is shown in section \ref{ssec:isolatedsphere} that this error has a minor effect. 




\subsubsection{Estimation of the desired velocity}
\label{ssec:desiredvelocity}
The general procedure to obtain the desired volume-filtered fluid velocity is to apply a defiltering operation to the volume-filtered fluid velocity at the Lagrangian surface marker, compute the deviation of the defiltered fluid velocity from the velocity of the particle (a no-slip boundary condition is enforced), and filter the deviation, which is then added to the volume-filtered fluid velocity at the Lagrangian surface marker. It is assumed that a defiltering operator $\mathcal{DF}$ can be constructed, such that
\begin{align}
    \mathcal{DF}\{ \overline{I_{\mathrm{f}}u_i}\vert_{\sigma_{\mathrm{L}}} \} = I_{\mathrm{f}} u_i.
\end{align}
Filtering with a Gaussian kernel is a reversible operation, which can be shown in Fourier space according to the convolution theorem
\begin{align}
    \hat{\Bar{\varPhi}} = \hat{g}\hat{\varPhi},
\end{align}
where $\hat{.}$ indicates the Fourier transform of a quantity. After rearranging, the Fourier transform of the unfiltered quantity is obtained as 
\begin{align}
    \hat{\varPhi} = \dfrac{\hat{\Bar{\varPhi}}}{\hat{g}}.
\end{align}
However, in the present framework the filtered quantities are not continuous functions but discrete values on a grid. The filtering of discrete data can only be inverted approximately and requires regularization to avoid amplification of numerical errors. In the present paper, we apply the approximate deconvolution method originally proposed by \citet{VanCittert1931} and later applied to LES by \citet{Stolz1999}, according to which
\begin{align}
\label{eq:approximatedeconvolution}
     \mathcal{DF}\{ \overline{I_{\mathrm{f}}u_i}\vert_{\sigma_{\mathrm{L}}} \} &\approx \sum_{p=0}^{N_{\mathrm{ad}}}(1 - g_{\mathrm{L}}\ast)^p\overline{I_{\mathrm{f}}u_i}\vert_{\sigma_L} \\ 
     &= \overline{I_{\mathrm{f}}u_i}\vert_{\sigma_L} + ( \overline{I_{\mathrm{f}}u_i}\vert_{\sigma_L} - g_{\mathrm{L}}\ast \overline{I_{\mathrm{f}}u_i}\vert_{\sigma_L}) + (1-2g_{\mathrm{L}}\ast\overline{I_{\mathrm{f}}u_i}\vert_{\sigma_L} + (g_{\mathrm{L}}\ast)^2\overline{I_{\mathrm{f}}u_i}\vert_{\sigma_L}) + \cdots,
\end{align}
where $\ast$ is the convolution operator. Note that the $p$-th power of $g\ast$ denotes $p$ applications of the convolution operation
\begin{align}
    (g\ast)^p\varPhi =  \underbrace{g\ast ( g\ast  (\cdots(g\ast\varPhi}_{p \text{ times}}) \cdots) ),
\end{align}
with the flow quantity $\varPhi$ and the filter kernel $g$. \\
If $\overline{I_{\mathrm{f}}u_i}\vert_{\sigma_L}$ is the exact filtered function and $N_{\mathrm{ad}}=\infty$, equation \eqref{eq:approximatedeconvolution} is exact. However, $\overline{I_{\mathrm{f}}u_i}\vert_{\sigma_L}$ is known discretely only, and, therefore, $N_{\mathrm{ad}}$ is finite. Hence, equation \eqref{eq:approximatedeconvolution} is an approximation in practice. For the LES framework, it is reported that increasing the number of approximate deconvolution levels beyond $N_{\mathrm{ad}}=5$ does not further improve the results \cite{Stolz2001}. \\
As becomes evident from equation \eqref{eq:approximatedeconvolution}, the approximate deconvolution requires multiple explicit filtering operations of $\overline{I_{\mathrm{f}}u_i}\vert_{\sigma_L}$ at the Lagrangian surface markers. A quantity at the Lagrangian surface marker can be obtained for an arbitrary filter width, $\sigma\geq \sigma_{\mathrm{L}}$, by interpolating the flow quantities on the Eulerian fluid mesh with a Gaussian interpolation kernel, $g_{\delta}$, with a suitable filter width, $\sigma_{\delta}$. \\
In Appendix \ref{ap:zerofilterwidthlimit}, it is shown that the volume-filtered fluid velocity converges to $\overline{I_{\mathrm{f}}u_i}\vert_{\sigma\rightarrow0} = u_{\mathrm{Int},l,i}/2$ in the limit of zero filter width, where $u_{\mathrm{Int},l,i}$ is the velocity of the interface at the Lagrangian surface marker with the index $l$. In the unfiltered space, the defiltered fluid velocity at the particle surface deviates by a small velocity, $\delta u_{l,i}$, from half the velocity of the interface due to discretization and modeling errors, which gives
\begin{align}
   \delta u_{l,i} =  \mathcal{DF}\{ \overline{I_{\mathrm{f}}u_i}\vert_{\sigma_{\mathrm{L}}}(\boldsymbol{X}_l) \} - \dfrac{1}{2}u_{\mathrm{Int},l,i}.
\end{align}
After rearranging and subsequent volume-filtering with $g_{\mathrm{L}}$, the desired volume-filtered fluid velocity is obtained as
\begin{align}
\label{eq:desiredfilteredvelocity}
(\overline{I_{\mathrm{f}}u_i}\vert_{\sigma_{\mathrm{L}}})_{\mathrm{des}}(\boldsymbol{X}_l) = \overline{I_{\mathrm{f}}u_i}\vert_{\sigma_{\mathrm{L}}}(\boldsymbol{X}_l) - \epsilon_{\mathrm{f}}\delta u_{l,i},
\end{align}
where $\delta u_{l,i}$ is assumed to be constant in the volume-filtering operation. Here, the fluid volume fraction $\epsilon_{\mathrm{f}}$ corresponds to a filter width $\sigma_{\mathrm{L}}$. 

\subsection{Implementation details of \newIBM{}}
\label{ssec:convertIBM}
Although the volume-filtering background of the \newIBM{} is relatively involved, only few relatively simple modifications are required to convert an implementation of a \oldIBM{} to a \newIBM{}. The general steps are outlined in the present section. \\
\subsubsection{Modification of the governing flow equations}
\label{ssec:modificationflowequation}
The viscous closure given by equation \eqref{eq:viscousclosure}, the subfilter stress closure given in equation \eqref{eq:subfilterstressclosure}, and source term in the continuity equation, $\partial \epsilon_{\mathrm{f}}/\partial t$, are added as explicit source terms to the continuity and fluid momentum equation. These additional terms contain the fluid- or particle volume fraction and spatial derivatives of these, corresponding to the filter width $\sigma_{\mathrm{E}}$, since these source terms are computed on the Eulerian fluid mesh. Since the total volume of the fluid and the total volume of the particles are both conserved, temporal derivatives of the volume fraction can be written as spatial derivatives of the volume fraction according to (see \citet{Hausmann2024a} for details)
\begin{align}
    \dfrac{\partial \epsilon_{\mathrm{f}}}{\partial t} = \sum\displaylimits_q v_{q,i}\dfrac{\partial \epsilon_{\mathrm{p},q}}{\partial x_i}.
\end{align}
The particle volume fraction for a Gaussian filter kernel of the particle with the index $q$ is given as \cite{Balachandar2022}
\begin{align}
\label{eq:particlevolumefraction}
    \epsilon_{\mathrm{p},q}(\boldsymbol{x}) = \dfrac{\mathrm{erf}(A)-\mathrm{erf}(B)}{2}+\dfrac{\exp(-A^2) - \exp(-B^2)}{\sqrt{\pi} (A+B)},
\end{align}
with 
\begin{align}
    A=\dfrac{2 |\boldsymbol{x}-\boldsymbol{x}_{\mathrm{p},q}|+d_{\mathrm{p},q}}{2\sqrt{2}\sigma}, \\
    B=\dfrac{2 |\boldsymbol{x}-\boldsymbol{x}_{\mathrm{p},q}|-d_{\mathrm{p},q}}{2\sqrt{2}\sigma},
\end{align}
with the particle diameter, $d_{\mathrm{p},q}$. \\
In a finite volume framework, the particle volume fraction and its spatial derivatives are required as integrals over the cell volume. In the present paper, the integrals of equation \eqref{eq:particlevolumefraction} and its spatial derivatives over the fluid mesh cells are approximated by the midpoint rule quadrature using four support points per direction. Since the particle volume fraction and its spatial derivatives rapidly decay with increasing distance from the particle center, the quadrature is only carried out over fluid mesh cells that satisfy $|\boldsymbol{x}_{m} - \boldsymbol{x}_{\mathrm{p},q}|<(4\sigma+0.5d_{\mathrm{p},q})$, where $\boldsymbol{x}_{m}$ is the center of a fluid mesh cell.  

\subsubsection{Interpolating the Eulerian fluid quantities to the Lagrangian surface markers}
The interpolation weight associated to each fluid mesh cell is obtained by analytical integration of the Gaussian over the fluid mesh cell. Since the Gaussian has an infinite support, the interpolation kernel is truncated after a distance of $4\sigma_{\delta}$ from the Lagrangian surface marker, where the Gaussian interpolation weight is essentially zero. If the standard deviation of the Gaussian interpolation kernel is chosen $\sigma_{\delta}=\sigma_{\mathrm{E}}$, these weights can be used for interpolation and spreading. 

\subsubsection{Determining the desired fluid velocity at the particle surface by approximate deconvolution} 
The computation of the desired volume-filtered fluid velocity requires defiltering. With approximate deconvolution, as given in equation \eqref{eq:approximatedeconvolution}, the defiltered fluid velocity is estimated by consecutive volume-filtering of the already volume-filtered fluid velocity. At the Lagrangian surface markers, the fluid velocity possesses a filter width $\sigma_{\mathrm{L}}$, since, as explained in section \ref{ssec:interpolationequalsspreading},
\begin{align}
    \overline{I_{\mathrm{f}}u_i}\vert_{\sigma_{\mathrm{L}}} = \overline{\overline{I_{\mathrm{f}}u_i}\vert_{\sigma_{\mathrm{E}}}}\vert_{\sigma_{\mathrm{E}}},
\end{align}
where it is assumed that the interpolation of fluid quantities and spreading of the momentum source is done with the same filter width, i.e., $\sigma_{\delta}=\sigma_{\mathrm{E}}$. According to equation \eqref{eq:subsequentfiltering}, subsequent  volume-filtering with the filter width $\sigma_{\mathrm{L}}$ gives
\begin{align}
    \overline{\overline{I_{\mathrm{f}}u_i}\vert_{\sigma_{\mathrm{L}}}}\vert_{\sigma_{\mathrm{L}}} = \overline{\overline{\overline{I_{\mathrm{f}}u_i}\vert_{\sigma_{\mathrm{E}}}}\vert_{\sigma_{\mathrm{E}}}}\vert_{\sqrt{2}\sigma_{\mathrm{E}}} = \overline{\overline{I_{\mathrm{f}}u_i}\vert_{\sigma_{\mathrm{E}}}}\vert_{\sqrt{3}\sigma_{\mathrm{E}}}.
\end{align}
Consequently, the repeatedly volume-filtered fluid velocity at the Lagrangian surface markers can be obtained by interpolating the Eulerian fluid velocity with a wider interpolation kernel, e.g., using a Gaussian interpolation kernel with a filter width $\sqrt{3}\sigma_{\mathrm{E}}$ to obtain $\overline{\overline{I_{\mathrm{f}}u_i}\vert_{\sigma_{\mathrm{L}}}}\vert_{\sigma_{\mathrm{L}}}$. With this the approximate deconvolution with $N_{\mathrm{ad}}=5$ and $\sigma_{\delta}=\sigma_{\mathrm{E}}$ is given as (see equation \eqref{eq:approximatedeconvolution})
\begin{align}
\label{eq:adm5}
    &\mathcal{DF}\{ \overline{I_{\mathrm{f}}u_i}\vert_{\sigma_{\mathrm{L}}} \} \approx 6\overline{I_{\mathrm{f}}u_i}\vert_{\sigma_{\mathrm{L}}} -15 \overline{I_{\mathrm{f}}u_i}\vert_{\sqrt{2}\sigma_{\mathrm{L}}} + 20\overline{I_{\mathrm{f}}u_i}\vert_{\sqrt{3}\sigma_{\mathrm{L}}} -15\overline{I_{\mathrm{f}}u_i}\vert_{\sqrt{4}\sigma_{\mathrm{L}}} + 6\overline{I_{\mathrm{f}}u_i}\vert_{\sqrt{5}\sigma_{\mathrm{L}}} -\overline{I_{\mathrm{f}}u_i}\vert_{\sqrt{6}\sigma_{\mathrm{L}}} \nonumber \\
    &= 6\overline{\overline{I_{\mathrm{f}}u_i}\vert_{\sigma_{\mathrm{E}}}}\vert_{\sigma_{\mathrm{E}}} - 15\overline{\overline{I_{\mathrm{f}}u_i}\vert_{\sigma_{\mathrm{E}}}}\vert_{\sqrt{3}\sigma_{\mathrm{E}}} + 20\overline{\overline{I_{\mathrm{f}}u_i}\vert_{\sigma_{\mathrm{E}}}}\vert_{\sqrt{5}\sigma_{\mathrm{E}}} -15\overline{\overline{I_{\mathrm{f}}u_i}\vert_{\sigma_{\mathrm{E}}}}\vert_{\sqrt{7}\sigma_{\mathrm{E}}}+6\overline{\overline{I_{\mathrm{f}}u_i}\vert_{\sigma_{\mathrm{E}}}}\vert_{\sqrt{9}\sigma_{\mathrm{E}}}-\overline{\overline{I_{\mathrm{f}}u_i}\vert_{\sigma_{\mathrm{E}}}}\vert_{\sqrt{11}\sigma_{\mathrm{E}}}.
\end{align}

\subsubsection{Computation and spreading of the fluid stress vector}
In a \oldIBM{}, the fluid momentum balance is evaluated at the particle surface to obtain a force per unit volume. Typically, a volume is associated to each Lagrangian point at the particle surface, which is sometimes taken equal to the volume of a fluid mesh cell \cite{Uhlmann2005,Tschisgale2017} or obtained by enforcing that spreading is the inversion of interpolation \cite{Pinelli2010,AbdolAzis2018}. Multiplying the force per unit volume with the volume associated to the Lagrangian point gives a force, which is spread to the Eulerian fluid mesh using different kernels that are constructed such that they satisfy certain desired properties (see, e.g., \citet{Peskin2003}). \\
Instead, the \newIBM{} solves the fluid momentum balance evaluated at the particle surface to obtain the stress vector, which has the dimension of a force per unit surface area. The assumption that the stress vector at the particle surface is approximately constant within the support of the filter kernel, which is given in equation \eqref{eq:stressconstantwithingsupport}, allows to define a length scale for each Lagrangian surface marker, $\mathcal{L}(\mathbf{X}_l)$, according to
\begin{align}
    \mathcal{L}(\mathbf{X}_l) = \dfrac{1}{\sum_{m=1}^{N_{\mathrm{s}}} g_{\mathrm{L}}(|\boldsymbol{X}_l-\boldsymbol{X}_m|) \Delta A_m}.
\end{align}
Note that in the case of spherical particles, the length scale, $\mathcal{L}(\mathbf{X}_l)$, can be evaluated analytically, which is detailed in Appendix \ref{ap:analyticallengthscale}. The expression for $\mathcal{L}(\mathbf{X}_l)$ is a direct consequence from volume filtering and equation \eqref{eq:stressconstantwithingsupport} and is introduced by \citet{Dave2023}. Therefore, the fluid stress vector is obtained as
\begin{align}
\label{eq:computationstressvector}
    \boldsymbol{\Tilde{\Sigma}}_{l} = \boldsymbol{b}_l \mathcal{L}(\boldsymbol{X}_l),
\end{align}
where no summation is carried out over the index $l$. The vector $\boldsymbol{b}_l$ contains the terms of the volume-filtered momentum equation interpolated to the Lagrangian surface marker that do not belong to the particle momentum source
\begin{align}
    b_{l,i} = -\rho_{\mathrm{f}}\dfrac{(\overline{I_{\mathrm{f}}u_i}\vert_{\sigma_{\mathrm{L}}})_{\mathrm{des}}(\boldsymbol{X}_l) - (\overline{I_{\mathrm{f}}u_i}\vert_{\sigma_{\mathrm{L}}})^n(\boldsymbol{X}_l)}{\Delta t} -\mathcal{A}_i^n(\boldsymbol{X}_l) - \mathcal{P}_i^n(\boldsymbol{X}_l) + \mathcal{V}_i^n(\boldsymbol{X}_l).
\end{align}
Once the fluid stress vector is computed, the particle momentum source equation \eqref{eq:particlemomentumsource} is discretized and computed as
\begin{align}
\label{eq:discretespreading}
     s_i(\boldsymbol{x}) \approx \sum_{l=1}^{N_{\mathrm{s}}} g_{\mathrm{E}}(|\boldsymbol{x}-\boldsymbol{X}_l|)\Tilde{\Sigma}_{l,i} \Delta A_l.
\end{align}

\subsubsection{One time step with the \newIBM{}}
One time step of the \newIBM{} can be summarized as follows:
\begin{enumerate}
    \item Moving of the particles and the Lagrangian surface markers using the fluid forces computed in the previous time step.
    \item Computation of the subfilter stress closure with equation \eqref{eq:subfilterstressclosure} and the viscous closure with equation \eqref{eq:viscousclosure}. The particle volume fraction and its spatial derivatives are computed using the quadrature described in section \ref{ssec:modificationflowequation}.
    \item Computation of the desired volume-filtered velocity at the particle surface according to equation \eqref{eq:desiredfilteredvelocity} using the approximate deconvolution given by equation \eqref{eq:adm5}.
    \item Interpolation of all terms occurring in the volume-filtered fluid momentum equation at the Lagrangian surface markers that are required to compute $\boldsymbol{b}_l$.
    \item Computation of the fluid stress vector at each Lagrangian surface marker according to equation \eqref{eq:computationstressvector} and the resulting fluid force on each particle.
    \item Spreading of the stress vector to the Eulerian fluid mesh by using the discretized particle momentum source as given in equation \eqref{eq:discretespreading}.
    \item Solution of the volume-filtered NSE \eqref{eq:reducedNSEconstinuity}-\eqref{eq:reducedNSEmomentum} with the subfilter stress closure, the viscous closure, and the particle momentum source. 
\end{enumerate}

%% file: Sections/results.tex
\begin{figure}[h]
    \centering
    \hspace{-1cm}
    \includegraphics[scale=0.75]{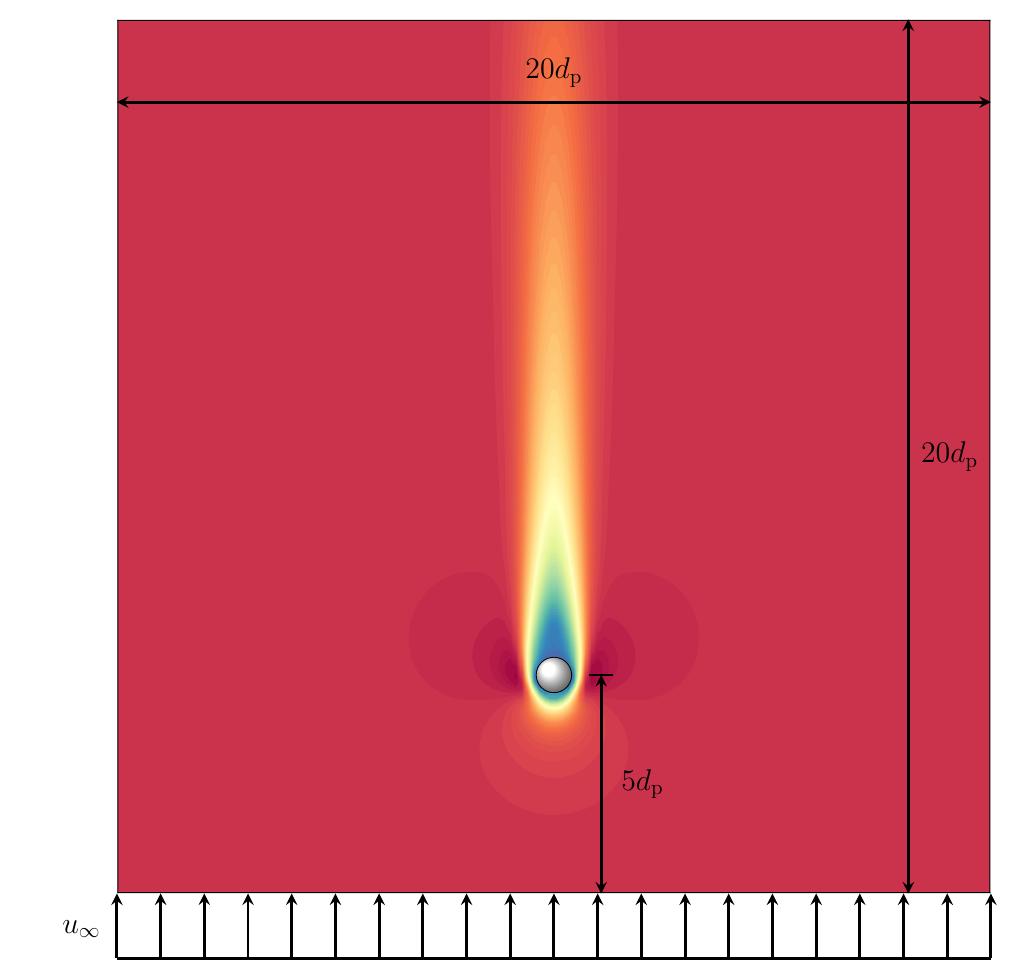}
    \caption{Sketch of the flow configuration for the flow over an isolated sphere. The particle is placed five particle diameters away from the inlet where a uniform inflow fluid velocity, $u_{\infty}$, is applied.}
    \label{fig:sketchisolatedsphere}
\end{figure}

\section{Validation, results, and discussion}
\label{sec:results}
In this section, the \newIBM{} is validated in different particle-laden flow configurations, such as the flow around an isolated spherical particle at different Reynolds numbers, the flow around arranged arrays of spheres, the flow around a dense random packing of spheres, the settling of a single sphere and the settling of many interacting spheres. In particular, we study the convergence properties of the \newIBM{}, the influence of the filter width, and the ability of the \newIBM{} to make accurate predictions with a coarse fluid mesh resolution. \\
If not stated differently, the simulations with the \newIBM{} have the following parameters: $\sigma_{\mathrm{E}}/\Delta x= 0.75$, $N_{\mathrm{ad}}=5$, $N_{\mathrm{s}}=1280$, and $\mathrm{CFL}=u_{\mathrm{max}} \Delta t/\Delta x\leq 0.05$, where $\Delta t$ is the simulation time step and $\Delta x$ is the spacing of the fluid mesh.

\subsection{Flow around a fixed isolated sphere}
\label{ssec:isolatedsphere}
First, we study the flow around a fixed isolated spherical particle as depicted in figure \ref{fig:sketchisolatedsphere}. The spherical particle with the diameter, $d_{\mathrm{p}}$, is placed $5d_{\mathrm{p}}$ away from the inlet in a cubic domain with the side length of $20d_{\mathrm{p}}$. At the outlet boundary, all normal fluid velocity gradients and the pressure are set to zero. At the inlet boundary and the remaining boundaries on the sides, the streamwise fluid velocity is $u_{\infty}$ and the other velocity components and the normal pressure gradient are set to zero. For the flow around an isolated sphere, the particle Reynolds number is defined as $\mathrm{Re}=\rho_{\mathrm{f}}u_{\infty}d_{\mathrm{p}}/\mu_{\mathrm{f}}$. \\

\begin{figure}[h]
    \centering
    \hspace{-2cm}
    \includegraphics{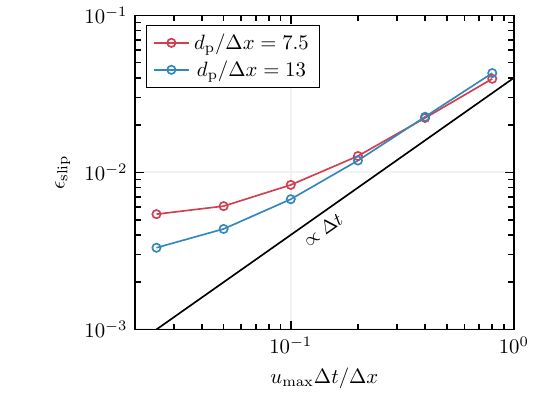}
    \caption{Temporal convergence of the maximum slip error at the particle surface for the flow around an isolated sphere at $\mathrm{Re}_{\mathrm{p}}=100$ and two resolutions, $d_{\mathrm{p}}/\Delta x=7.5$ and $d_{\mathrm{p}}/\Delta x=13$. The time step is normalized by the fluid mesh spacing and the maximum fluid velocity in the flow domain, $u_{\mathrm{max}}$.}
    \label{fig:temporalconvergence}
\end{figure}
We study the ability of the \newIBM{} to enforce the desired fluid velocity at the particle surface. The following errors may cause a deviation of the fluid velocity at the particle surface from the desired velocity:
\begin{itemize}
    \item the discretization error of the time derivative in equation \eqref{eq:momentumatsurface},
    \item the assumption that the stress vector is approximately constant within the support of $g_{\mathrm{L}}$, i.e., equation \eqref{eq:stressconstantwithingsupport},
    \item discretization errors of the volume-filtered NSE on the Eulerian fluid mesh,
    \item errors due to truncation of the support of the Gaussian interpolation and spreading kernel.
\end{itemize}
The deviation from the desired fluid velocity is quantified by means of the maximum slip error, which is defined as
\begin{align}
    \epsilon_{\mathrm{slip}} = \max (|(\epsilon_{\mathrm{f}}\Bar{u}_i)_{\mathrm{des}} - \epsilon_{\mathrm{f}}\Bar{u}_i|)/u_{\infty},
\end{align}
where the maximum operator is with respect to every Lagrangian surface marker. Figure \ref{fig:temporalconvergence} shows $\epsilon_{\mathrm{slip}}$ as a function of the normalized time step for the case of the flow around an isolated sphere at $\mathrm{Re}_{\mathrm{p}}=100$. Two spatial resolutions are shown in figure \ref{fig:temporalconvergence}, $d_{\mathrm{p}}/\Delta x=7.5$ and $d_{\mathrm{p}}/\Delta x=13$. It is observed that $\epsilon_{\mathrm{slip}}$ grows linearly with the time step for non-dimensional time steps $u_{\mathrm{max}} \Delta t/\Delta x>0.2$ for both spatial resolutions, which suggests that the discretization error of the temporal derivative in equation \eqref{eq:momentumatsurface} is dominant at these time steps as this is the only error linearly proportional to the time step. At very small time steps, the error approaches a plateau, suggesting that one or multiple errors take over, which are functions of the spatial discretization or the filter width, but not the time step. The presumably dominating error is the assumption that the stress vector is constant within the support of $g_{\mathrm{L}}$ represented by equation \eqref{eq:stressconstantwithingsupport}. With decreasing filter width this assumption is less restrictive and the associated slip error is smaller. At the higher spatial resolution, $d_{\mathrm{p}}/\Delta x=13$, the slip error contribution that is independent of the time step is significantly smaller than with the coarse spatial resolution, $d_{\mathrm{p}}/\Delta x=7.5$. At the higher spatial resolution, the smaller filter width reduces the error caused by assuming that the stress vector is constant within the support of $g_{\mathrm{L}}$ and the smaller $\Delta x$ reduces the discretization error at the Eulerian fluid mesh. Even for the coarse spatial resolution, however, a very small slip error can be achieved, which justifies the assumption introduced with equation \eqref{eq:stressconstantwithingsupport}. \\

\begin{figure}[h]
    \centering
    \hspace{-2cm}
    \includegraphics{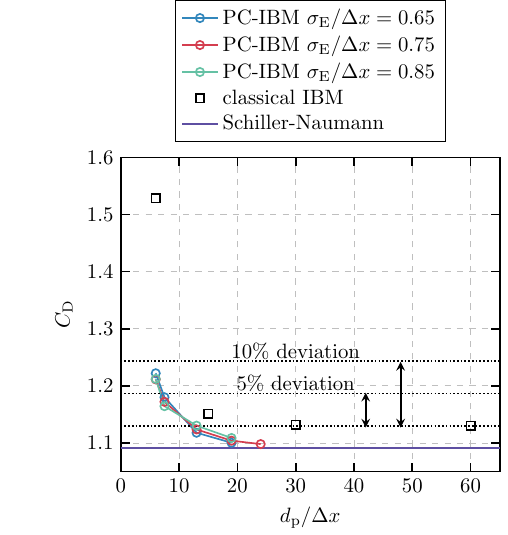}
    \caption{Convergence of the drag coefficient, $C_{\mathrm{D}}$, at $\mathrm{Re}_{\mathrm{p}}=100$ as a function of the spatial resolution. The obtained drag coefficients with the \newIBM{} are shown for different filter widths and compared to the \oldIBM{} with spatial resolutions between $d_{\mathrm{p}}/\Delta x=6$ and $d_{\mathrm{p}}/\Delta x=60$ and the drag coefficient obtained from the Schiller-Naumann correlation \cite{Schiller1933}. 5\% and 10\% deviation from the highly resolved \oldIBM{} are highlighted for comparison.}
    \label{fig:CDconvergence}
\end{figure}
\begin{figure}
    \centering
    \hspace{-1.75cm}
    \includegraphics{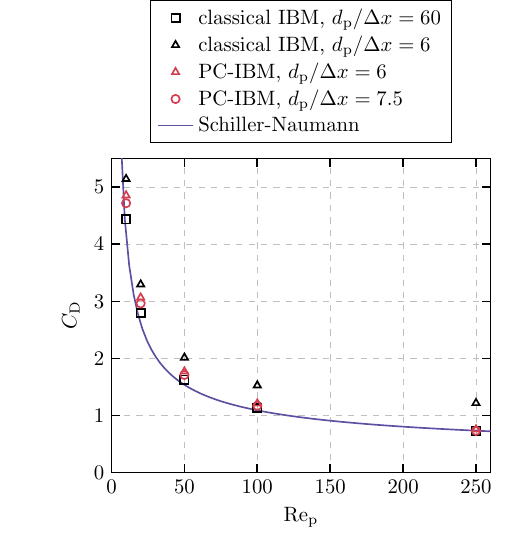}
    \caption{Drag coefficients predicted with the \newIBM{} and the \oldIBM{} for different Reynolds numbers and spatial resolutions. The Schiller-Naumann correlation is shown as reference \cite{Schiller1933}.}
    \label{fig:CD}
\end{figure}
\begin{figure}
    \centering
    \includegraphics{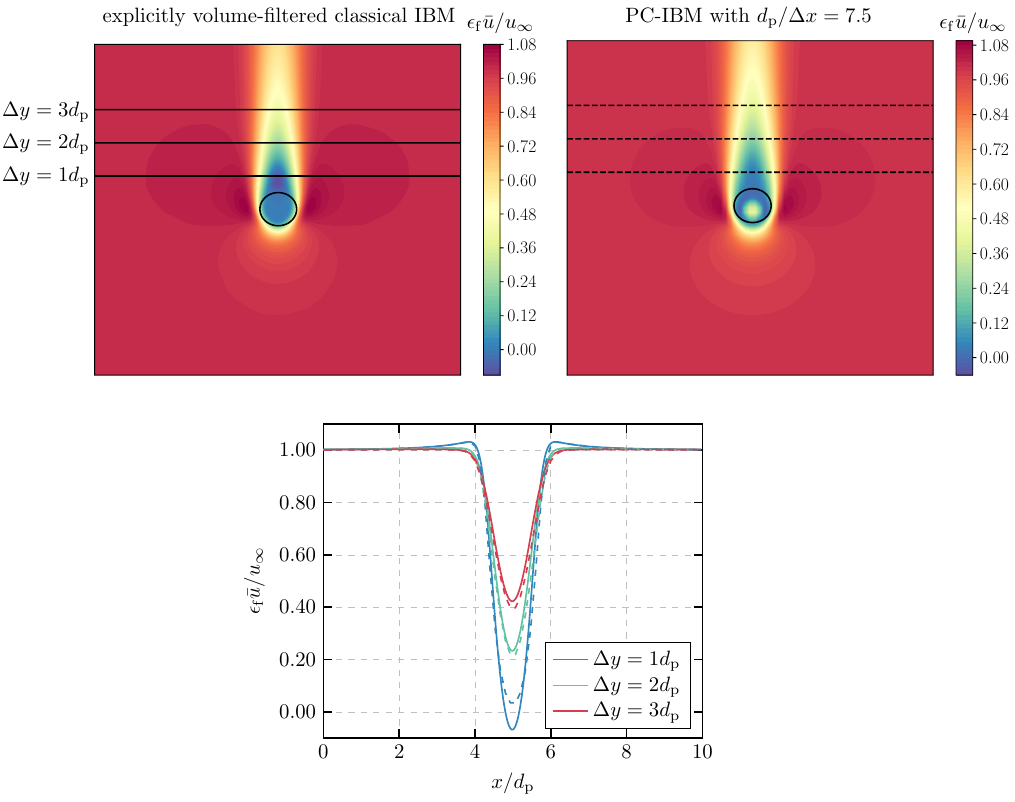}
    \caption{Comparison of the explicitly volume-filtered highly resolved \oldIBM{} and the \newIBM{} with $d_{\mathrm{p}}/\Delta x=7.5$ for the flow around an isolated sphere at $\mathrm{Re}_{\mathrm{p}}=100$. The volume-filtered fluid velocity is separately plotted along the indicated lines at different locations downstream of the particle, i.e., $\Delta y=1d_{\mathrm{p}}$,  $\Delta y=2d_{\mathrm{p}}$, and  $\Delta y=3d_{\mathrm{p}}$ downstream from the center of the particle. The solid lines correspond to the explicitly volume-filtered \oldIBM{} and the dashed lines to the \newIBM{}.}
    \label{fig:velocityfieldisolatedsphere}
\end{figure}

Figure \ref{fig:CDconvergence} shows the convergence of the drag coefficient, $C_{\mathrm{D}}$, as a function of the spatial resolution for $\mathrm{Re}_{\mathrm{p}}=100$, where the drag coefficient is defined as 
\begin{align}
    C_{\mathrm{D}} = \dfrac{F_{\mathrm{D}}}{\dfrac{\rho_{\mathrm{f}}}{2} u_{\infty}^2 \dfrac{\pi}{4} d_{\mathrm{p}}^2},
\end{align}
with the drag force, $F_{\mathrm{D}}$. The drag convergence of the \newIBM{} is shown for three different filter width to mesh spacing ratios, $\sigma_{\mathrm{E}}/\Delta x$, and compared against the \oldIBM{} and the Schiller-Naumann correlation \cite{Schiller1933}. For all of the three $\sigma_{\mathrm{E}}/\Delta x$, the drag coefficient converges towards the drag coefficient from the Schiller-Naumann correlation as the mesh resolution is increased, which slightly deviates from the drag coefficient obtained from the \oldIBM{} for the high resolution of $d_{\mathrm{p}}/\Delta x=60$. Note that it is not clear which drag coefficient is correct, but the drag coefficients obtained from the Schiller-Naumann correlation are typically considered as reference values (see, e.g., \citet{Dave2023}) and have been confirmed experimentally \cite{Johnson1999}. It can also be observed in figure \ref{fig:CDconvergence} that the influence of the filter width is insignificant. Generally, small ratios of $\sigma_{\mathrm{E}}/\Delta x$ may lead to spatial gradients which are too sharp to be resolved by the fluid mesh and result in a large discretization error, whereas a large ratio of $\sigma_{\mathrm{E}}/\Delta x$ generally increases the modeling errors, i.e., the errors of the modeled subfilter stress tensor, the approximate deconvolution, and the error associated to equation \eqref{eq:stressconstantwithingsupport}. We consider the ratio $\sigma_{\mathrm{E}}/\Delta x=0.75$ as a good trade-off between the two error sources. \\
It is worth noting, that, as observed in figure \ref{fig:CDconvergence}, the \newIBM{} predicts a drag coefficient which is still relatively accurate, even at very low resolutions. With $\sigma_{\mathrm{E}}/\Delta x=0.75$ and a spatial resolution of $d_{\mathrm{p}}/\Delta x=6$, the deviation of the drag coefficient compared to the highly resolved \oldIBM{} is less than 10\% and with a resolution of $d_{\mathrm{p}}/\Delta x=7.5$ the deviation is less than 5\%. The \oldIBM{} predicts a drag coefficient with around 40\% deviation with a spatial resolution of $d_{\mathrm{p}}/\Delta x=6$. \\
Figure \ref{fig:CD} shows the drag coefficients predicted by the \newIBM{} and the \oldIBM{} for a range of Reynolds numbers together with the Schiller-Naumann correlation. It can be observed that the \newIBM{} predicts the drag coefficient accurately for the whole range of investigated Reynolds numbers. The \newIBM{} with $d_{\mathrm{p}}/\Delta x=6$ never exceeds a 10\% deviation from the drag coefficient of the \oldIBM{} with a spatial resolution of $d_{\mathrm{p}}/\Delta x=60$. Especially at higher Reynolds numbers, the \newIBM{} accurately predicts the drag coefficient at resolutions where a \oldIBM{} shows large deviations of the predicted drag coefficient. \\
In order to assess the flow field generated by the \newIBM{}, the fluid velocity field of the highly resolved \oldIBM{} of the flow around an isolated particle at $\mathrm{Re}_{\mathrm{p}}=100$ is explicitly volume-filtered and compared to the fluid velocity field generated by the \newIBM{} of the same configuration with a resolution $d_{\mathrm{p}}/\Delta x=7.5$. Since the \newIBM{} solves the volume-filtered NSE \eqref{eq:reducedNSEconstinuity}-\eqref{eq:reducedNSEmomentum}, the resulting flow field of the \newIBM{} is interpreted as the volume-filtered fluid velocity field. Figure \ref{fig:velocityfieldisolatedsphere} shows the contours of the explicitly volume-filtered fluid velocity from the highly resolved \oldIBM{} and the \newIBM{}. At three different location in the wake of the particle, i.e., $\Delta y=1d_{\mathrm{p}}$,  $\Delta y=2d_{\mathrm{p}}$, and  $\Delta y=3d_{\mathrm{p}}$ downstream from the center of the particle, lines of the volume-filtered fluid velocity are extracted and plotted separately. For the lines at $\Delta y=2d_{\mathrm{p}}$ and $\Delta y=3d_{\mathrm{p}}$ downstream from the particle center, the volume-filtered fluid velocity exhibits an almost perfect match. Along the line at $\Delta y=1d_{\mathrm{p}}$ downstream from the particle center, small differences in the volume-filtered fluid velocity are observed with a maximum error of approximately 10\% of $u_{\infty}$. The deviation from the explicitly volume-filtered fluid velocity is located close to the particle but decreases in the spanwise directions with increasing distance to the particle. \\
Note that, although the volume-filtered fluid velocity of the \newIBM{} is in good agreement with explicitly volume-filtered fluid velocity outside the particle, significant deviations are observed inside the particle. Any volume-filtered quantity should be close to zero in the center of the particle for small filter widths, since the fluid volume fraction is very small. The deviation of volume-filtered quantities from zero inside the particle for small filter widths can be explained by the presence of the error of the enforced desired volume-filtered fluid velocity and the modeling error from the subfilter stress tensor. In the \newIBM{}, the desired velocity itself is an estimation from the truncated approximate deconvolution. Another possible explanation is that no boundary conditions of the volume-filtered pressure or pressure gradient are applied at the particle surface. However, the fluid velocities obtained from simulations with the \oldIBM{} outside the immersed boundary are typically in good agreement with experimental observations \cite{Mittal2005}, suggesting that the impact of the missing pressure boundary condition and the error made when enforcing the desired fluid velocity is more impactful inside the particle than outside the particle. 

\FloatBarrier

\subsection{Flow around arrays of spheres}
\label{ssec:arrayofspheres}
\begin{figure}
    \centering
    \includegraphics[scale=1]{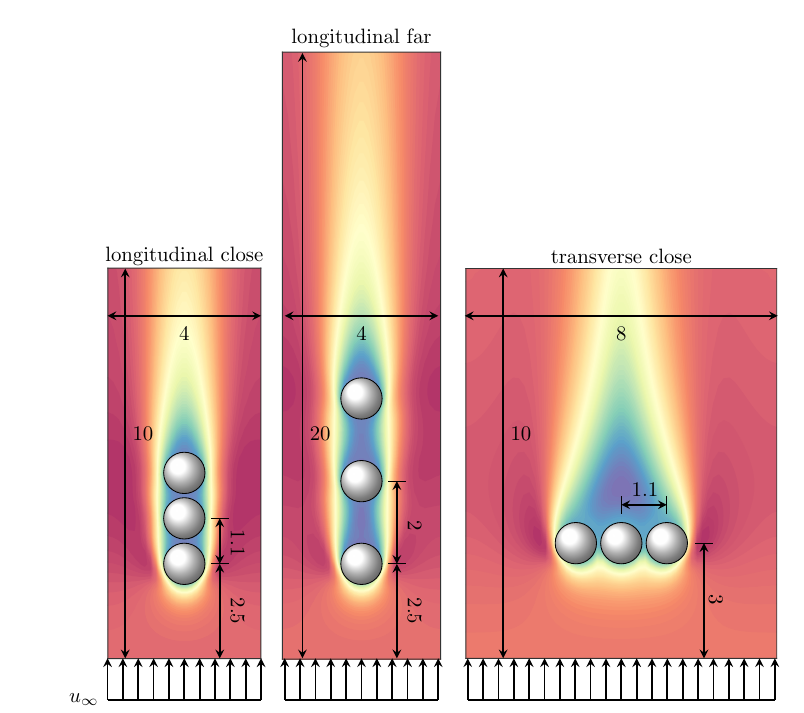}
    \caption{Sketch of the three simulations of ordered arrays of spheres. The given length scales are multiples of the particle diameter, $d_{\mathrm{p}}$, and all domains have a depths of $4d_{\mathrm{p}}$. In all three simulations, the inlet is a uniform flow of a velocity $u_{\infty}$. }
    \label{fig:sketcharrays}
\end{figure}

\begin{figure}
    \centering
    \hspace{-1.75cm}
    \includegraphics{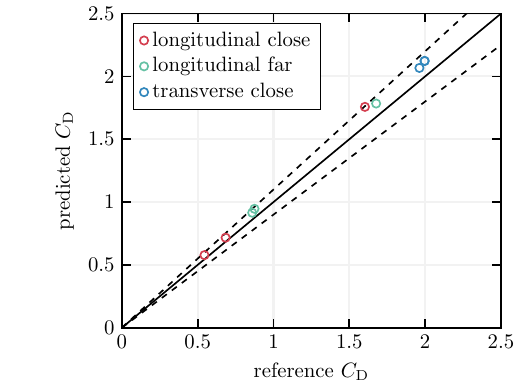}
    \caption{Drag coefficient predicted by the \newIBM{} with a resolution of $d_{\mathrm{p}}/\Delta x=7.5$ ($d_{\mathrm{p}}/\Delta x=8$ for the longitudinal far case) compared to the reference drag coefficient obtained from the \oldIBM{} (indicated as reference $C_{\mathrm{D}}$) with a resolution of $d_{\mathrm{p}}/\Delta x=60$. The dashed lines indicate a 10\% deviation from the actual value.}
    \label{fig:predictedforcearray}
\end{figure}

The flow around differently arranged arrays of spheres is investigated to assess the ability of the \newIBM{} to predict accurate forces on the particles when other particles are in the close vicinity. Three different particle arrangements, each consisting of three spherical particles, are considered and visualized in figure \ref{fig:sketcharrays}, a close longitudinal arrangement, a far longitudinal arrangement, and a close transverse arrangement. All three arrangements possess a uniform inflow velocity $u_{\infty}$, an outflow with zero velocity gradients, and symmetric boundary conditions at the sides with a zero normal velocity. Based on the uniform inflow velocity and the particle diameter, the particle Reynolds number of the three configurations is $\mathrm{Re}_{\mathrm{p}}=50$. \\
The \newIBM{} simulations are carried out with a resolution of $d_{\mathrm{p}}/\Delta x=7.5$ or $d_{\mathrm{p}}/\Delta x=8$, and the resulting forces on the particles are compared to similar simulations with the \oldIBM{} and a resolution of $d_{\mathrm{p}}/\Delta x=60$. The forces on the particles in the streamwise direction are normalized and represented using the drag coefficient that is calculated with the uniform inflow velocity, $u_{\infty}$. In figure \ref{fig:predictedforcearray}, the resulting drag coefficients of the \newIBM{} with $d_{\mathrm{p}}/\Delta x=7.5$ and $d_{\mathrm{p}}/\Delta x=8$ are compared with the drag coefficients obtained from the \oldIBM{} with $d_{\mathrm{p}}/\Delta x=60$. It can be observed that none of the predicted drag coefficients exceeds a 10\% deviation from the reference values.

\FloatBarrier

\begin{figure}[h]
    \centering
    \includegraphics[scale=0.1]{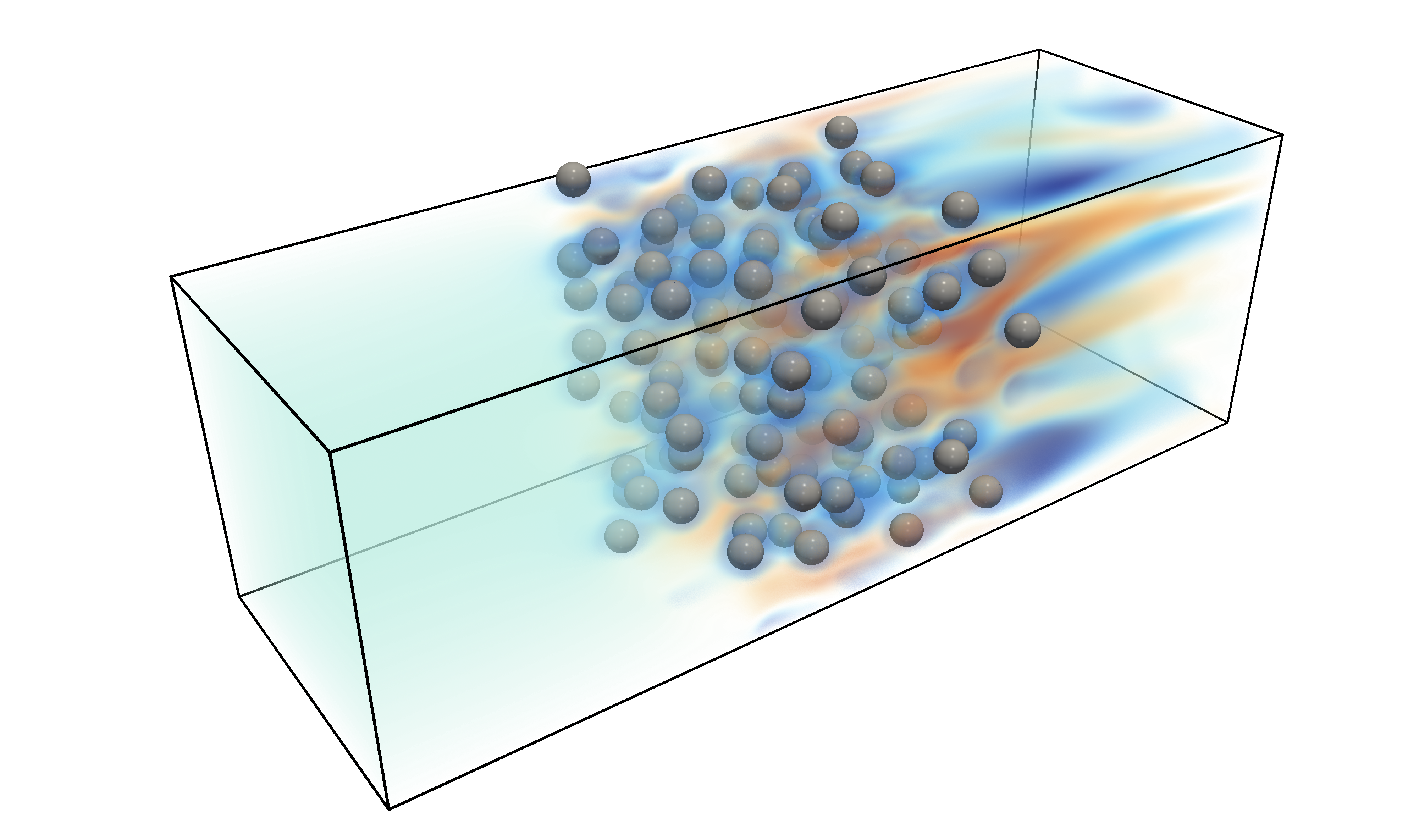}
    \caption{Sketch of the simulation configuration of the dense packing of spheres.}
    \label{fig:sketchdensepacking}
\end{figure}
\subsection{Dense packing of spheres}
\begin{figure}[h]
    \centering
    \hspace{-1.5cm}
    \includegraphics{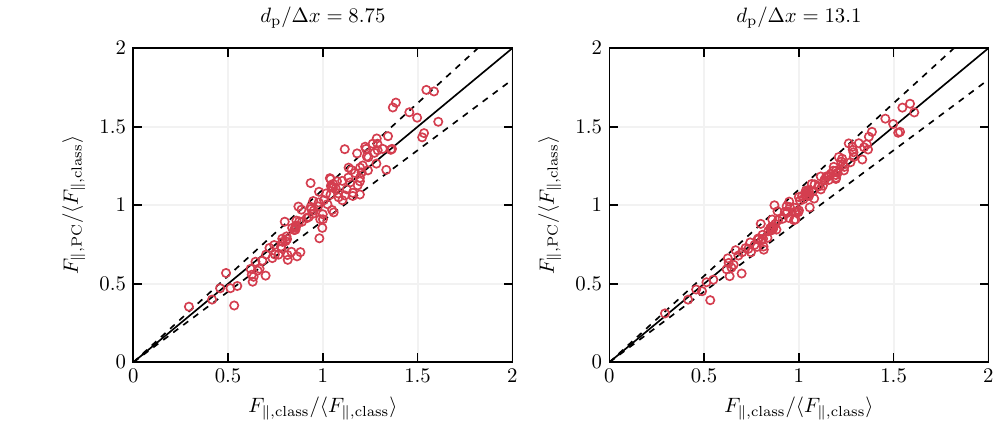}
    \caption{Normalized streamwise force predicted by the \newIBM{} with a resolution of $d_{\mathrm{p}}/\Delta x=8.75$ (left) and $d_{\mathrm{p}}/\Delta x=13.1$ (right) compared to the actual normalized streamwise force obtained from the \oldIBM{} with a resolution of $d_{\mathrm{p}}/\Delta x=30$. The dashed lines indicate 10\% deviation from the reference value.}
    \label{fig:forcesdensepacking}
\end{figure}

A complex flow configuration is generated by packing 123 spheres close together, such that a cube with a average particle volume fraction of $\langle \epsilon_{\mathrm{p}} \rangle = 0.2$ is obtained. Upstream and downstream of the particle-laden cube, cubes without particles are appended and a pressure drop from the inlet to the outlet, $\Delta p$, is applied to drive the flow. In addition to the inlet and the outlet, periodic boundary conditions are applied at the remaining boundaries. With the superficial velocity in the steady state, $u_{\mathrm{sf}}=\langle I_{\mathrm{f}}u_{\parallel} \rangle$, the Reynolds number of the present configuration is $\mathrm{Re}_{\mathrm{p}}=u_{\mathrm{sf}}d_{\mathrm{p}}\rho_{\mathrm{f}}/\mu_{\mathrm{f}}=110$, where the streamwise direction is indicated with the symbol $\parallel$. The computational domain has a size of $20.562 d_{\mathrm{p}}$ in the streamwise and $6.854 d_{\mathrm{p}}$ in the two other directions and is sketched in figure \ref{fig:sketchdensepacking}. \\
We perform a reference simulation with the \oldIBM{} with a resolution of $d_{\mathrm{p}}/\Delta x=30$. Simulations with the \newIBM{} are performed with two different resolutions, $d_{\mathrm{p}}/\Delta x=8.75$ and $d_{\mathrm{p}}/\Delta x=13.1$. The number of Lagrangian surface markers per particle is $N_{\mathrm{s}}=320$ for the dense packing configuration. As shown in table \ref{tab:densepacking}, the mean streamwise force on the particles obtained using the \newIBM{}, $\langle F_{\parallel,\mathrm{PC}} \rangle$, is nearly identical to the mean streamwise force obtained with the highly resolved \oldIBM{}, $\langle F_{\parallel,\mathrm{class}}\rangle$. The superficial velocity obtained with the \newIBM{}, $u_{\mathrm{sf,PC}}=\langle \epsilon_{\mathrm{f}} \Bar{u}_{\parallel} \rangle$, is smaller than the superficial velocity obtained from the highly resolved \oldIBM{}. With the \newIBM{} and a spatial resolution of $d_{\mathrm{p}}/\Delta x=13.1$, the deviation is approximately 3.2\%. \\
\begin{table}[h]
    \centering
     \caption{Comparison of the mean streamwise force on the particles and the superficial velocity obtained with two coarse resolution using the \newIBM{} with the values obtained using the highly resolved \oldIBM{}.}
    \label{tab:densepacking}
    \begin{tabular}{ccc}
    \toprule
    Resolution & $d_{\mathrm{p}}/\Delta x=8.75$ & $d_{\mathrm{p}}/\Delta x=13.1$  \\
    \midrule
    $\langle F_{\parallel,\mathrm{PC}} \rangle/\langle F_{\parallel,\mathrm{class}}\rangle$ & $1.0073$ & $1.0091$ \\
    \midrule
    $u_{\mathrm{sf,PC}}/u_{\mathrm{sf}}$ & $0.8845$ & $0.9678$ \\
    \bottomrule
    \end{tabular}
\end{table}
In figure \ref{fig:forcesdensepacking}, the streamwise force on every individual particle obtained with the \newIBM{} is compared to the forces on the particles with the \oldIBM{}. For the resolution $d_{\mathrm{p}}/\Delta x=8.75$, the forces on the particles are scattered around the expected value, whereas approximately 28\% of the forces deviate more than 10\% from the force obtained from the highly resolved \oldIBM{}. Increasing the resolution of the \newIBM{} to $d_{\mathrm{p}}/\Delta x=13.1$ reduces the scattering of the forces around the expected force, and less than 6\% of the particles possess a streamwise force that deviates more than 10\% from the expected force. \\

The flow configuration of packed spheres with a particle volume fraction of $\langle \epsilon_{\mathrm{p}} \rangle = 0.2$ demonstrates that with increasing particle volume fraction, the resolution of the \newIBM{} has to be increased to predict the forces on the particles with a similar accuracy as in the isolated sphere configuration discussed in section \ref{ssec:isolatedsphere} or the configurations of arranged arrays of spheres that is discussed in section \ref{ssec:arrayofspheres}.  

\FloatBarrier

\subsection{Isolated settling sphere in a quiescent fluid}
\begin{figure}
    \centering
    \hspace{-2.5cm}
    \includegraphics{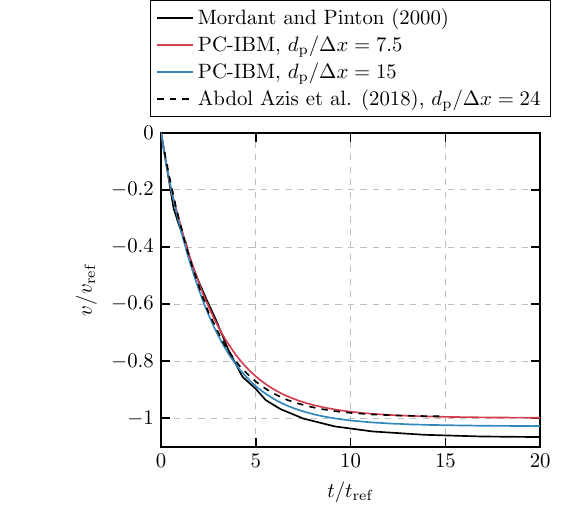}
    \caption{Normalized particle settling velocity over normalized time for the isolated settling sphere in quiescent fluid. The \newIBM{} with two resolutions is compared to the corresponding experiment of \citet{Mordant2000} and a \oldIBM{} with high resolution extracted from \citet{AbdolAzis2018}.}
    \label{fig:isolatedsettling}
\end{figure}

As discussed in section \ref{ssec:volumefilteredequations}, several contributions to the closures of the volume-filtered NSE vanish when the particle velocity is zero in the considered frame of reference. To demonstrate the accuracy of the implementation of the closure term contributions that arise when particles move, simulations of a configuration with a settling sphere in quiescent fluid are conducted. Note that the validity of the additional closure terms itself has been provided in a previous study \cite{Hausmann2024a}.\\
The considered simulation configuration involves a sphere with a diameter $d_{\mathrm{p}}$ settling under vertical gravitational acceleration, $g$, in a container with a vertical size, $L_{\mathrm{v}}$, and horizontal sizes, $L_{\mathrm{h}}$. Symmetric boundary conditions are applied at all boundaries, i.e, zero normal velocity, zero tangential velocity gradients, and zero normal pressure gradient. The sphere is released from a height $h_0$ and the particle Reynolds number, $\mathrm{Re}_{\mathrm{p}}$, is computed based on the terminal velocity of the particle. An experiment of a similar configuration from \citet{Mordant2000} is taken as reference for the present simulations. The parameters of the conducted simulations are listed in table \ref{tab:parameterssinglesedimentation}. \\
\begin{table}[h]
    \centering
     \caption{List of the simulation parameters for the isolated settling sphere in quiescent fluid.}
    \label{tab:parameterssinglesedimentation}
    \begin{tabular}{ccccc}
    \toprule
     $\mathrm{Re}_{\mathrm{p}}$ & $\rho_{\mathrm{p}}/\rho_{\mathrm{f}}$ & $L_{\mathrm{h}}/d_{\mathrm{p}}$ & $L_{\mathrm{v}}/d_{\mathrm{p}}$ & $h_0/d_{\mathrm{p}}$ \\
    \midrule
     $41$ & $2.56$ & $8$ & $30$ & $23$ \\
    \bottomrule
    \end{tabular}
\end{table}
In figure \ref{fig:isolatedsettling}, the particle settling velocity normalized by the reference velocity, $v_{\mathrm{ref}}=\sqrt{d_{\mathrm{p}}g}$, is shown over time normalized by the reference time, $t_{\mathrm{ref}}=\sqrt{d_{\mathrm{p}}/g}$, for simulations with the \newIBM{} with two spatial resolutions, $d_{\mathrm{p}}/\Delta x=7.5$ and $d_{\mathrm{p}}/\Delta x=15$. The time step is chosen such that $\mathrm{CFL}=u_{\mathrm{max}} \Delta t/\Delta x\leq 0.025$. The particle settling velocities are compared to the experiment of \citet{Mordant2000} and the \oldIBM{} with a resolution of $d_{\mathrm{p}}/\Delta x=24$ as reported in \citet{AbdolAzis2018}. The \newIBM{} with both resolutions slightly underestimates the magnitude of the terminal velocity, whereas the simulation with the higher resolution, $d_{\mathrm{p}}/\Delta x=15$, leads to a significantly smaller deviation from the terminal velocity observed in the experiment. However, even with the coarse resolution of $d_{\mathrm{p}}/\Delta x=7.5$, the deviation from the terminal velocity of the experiment is only approximately 6\%. The \oldIBM{} with a resolution of $d_{\mathrm{p}}/\Delta x=24$, as reported in \citet{AbdolAzis2018}, leads to approximately the same terminal velocity as the \newIBM{} with a spatial resolution of $d_{\mathrm{p}}/\Delta x=7.5$. \\
Since the closures of the volume-filtered NSE also occur in the fluid momentum equation evaluated at the particle surface, i.e., equation \eqref{eq:momentumatsurface}, the fluid force on the particle depends on the accurate implementation of the closures. Additionally, the closures affect the local volume-filtered fluid velocity and, therefore, the velocity seen by the particle. Since the isolated settling sphere configuration leads to results approximately as accurate as the isolated fixed sphere configuration, we infer that the quadrature of the volume fraction and its spatial derivatives as described in section \ref{ssec:convertIBM} is sufficiently accurate. 
\FloatBarrier
\begin{figure}
    \centering
    \includegraphics[scale=0.15]{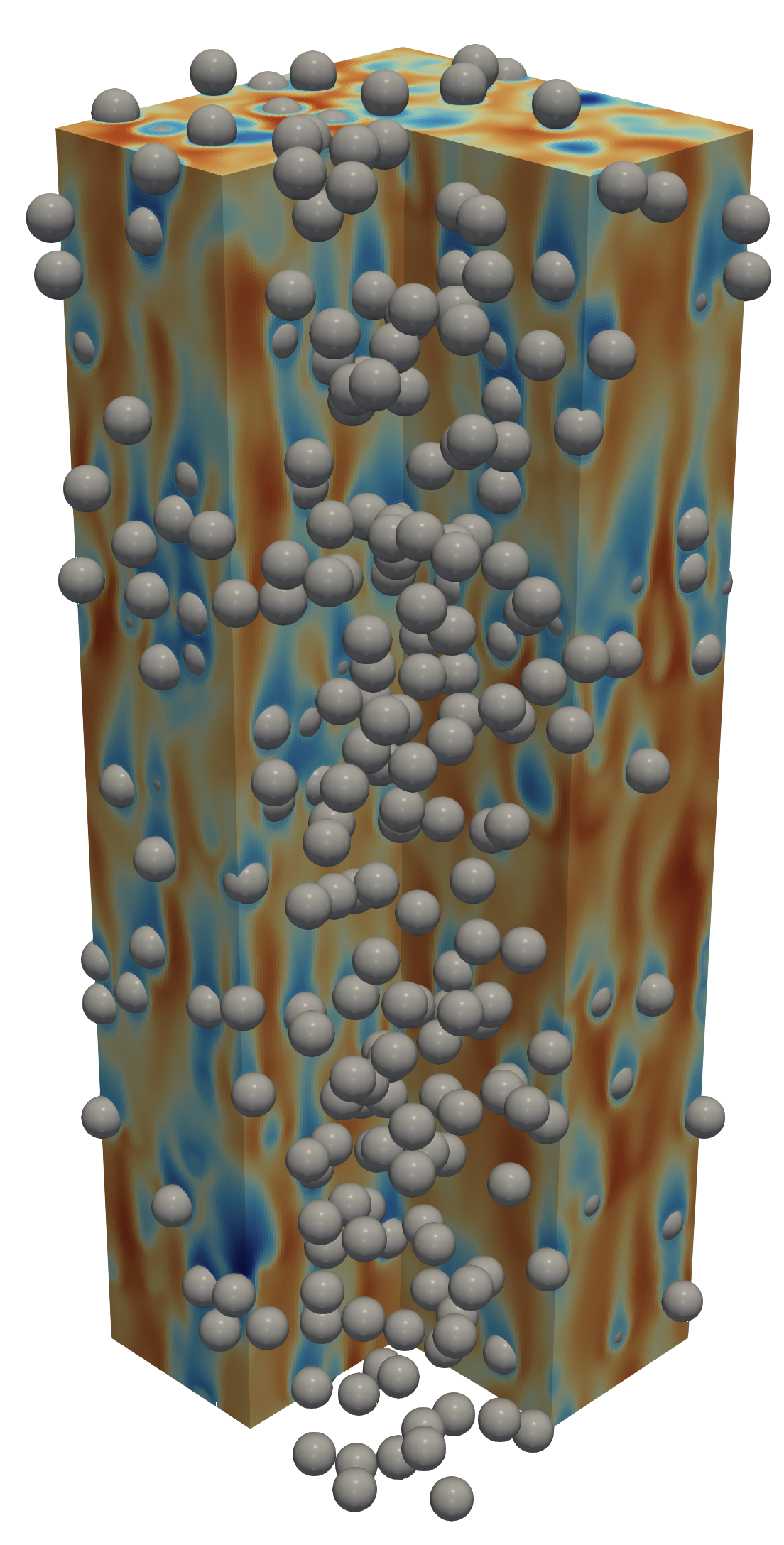}
    \caption{Sketch of the periodic settling of spheres.}
    \label{fig:sketchperiodicsettling}
\end{figure}
\subsection{Periodic settling of spheres}
\begin{figure}
    \centering
    \hspace{-2cm}
    \includegraphics{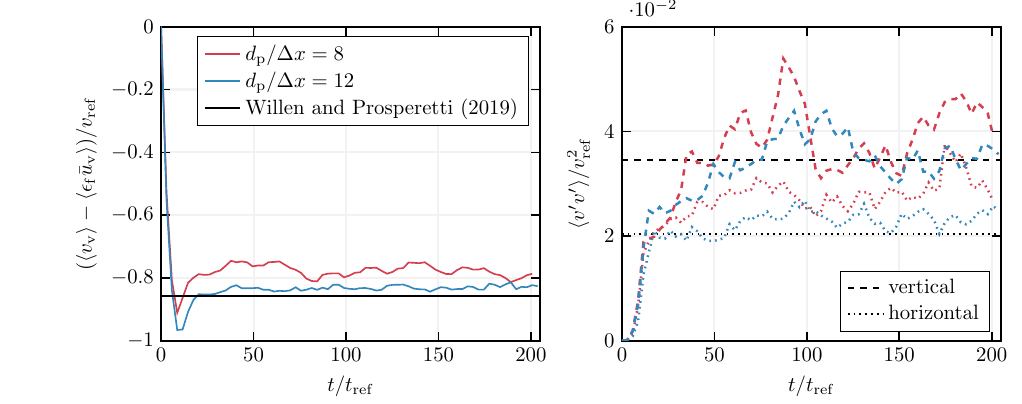}
    \caption{Mean relative vertical velocity between fluid and particles (left) and vertical and horizontal correlation of the particle velocity fluctuations (right) in the periodic settling configuration for the \newIBM{} with the resolutions $d_{\mathrm{p}}/\Delta x=8$ (red) and $d_{\mathrm{p}}/\Delta x=12$ (blue) compared to the averaged results reported in \citet{Willen2019}.}
    \label{fig:periodicsettlingresults}
\end{figure}

The final validation configuration studied in the present paper is the sedimentation of many spheres in a fully periodic domain as sketched in figure \ref{fig:sketchperiodicsettling}, which represents a complex application with practical relevance, e.g., to study the settling behavior of particles \cite{Uhlmann2014,Yin2007,Nicolai1995,Zaidi2015}. We conduct simulations with the \newIBM{} of a similar configuration as \citet{Willen2019}, who perform simulations using the Physalis method to couple the particles with the flow. \\
Since the configuration contains multiple moving spheres, collisions between particles may occur that require additional treatment to avoid an unphysical overlap of particles. In the present study, we apply the repulsive force model proposed by \citet{Glowinski2001}, with the adaptions reported in \citet{AbdolAzis2018}, that extend the range of applicable density ratios. Assuming that two spherical particles with the index $q$ and $r$ and the same size collide, the collision force on the particle with the index $q$ is computed as follows in the present simulations
\begin{align}
    F_{\mathrm{c},q,i} = 
    \begin{cases}
    0 &\text{if $\delta_{\mathrm{c}}>d_{\mathrm{p}}+\Delta r_{\mathrm{c}}$}\\
    \kappa_{\mathrm{c}}\left(\dfrac{d_{\mathrm{p}}+\Delta r_{\mathrm{c}} - \delta_{\mathrm{c}}}{\Delta r_{\mathrm{c}}} \right)^2 \dfrac{x_{\mathrm{p},q,i}-x_{\mathrm{p},r,i}}{\delta_{\mathrm{c}}}&\text{if $d_{\mathrm{p}}<\delta_{\mathrm{c}}\leq d_{\mathrm{p}}+\Delta r_{\mathrm{c}}$} \\
    \kappa_{\mathrm{c}}\left(\dfrac{d_{\mathrm{p}}+\Delta r_{\mathrm{c}} - \delta_{\mathrm{c}}}{\Delta r_{\mathrm{c}}} \right)^4 \dfrac{x_{\mathrm{p},q,i}-x_{\mathrm{p},r,i}}{\delta_{\mathrm{c}}}&\text{if $\delta_{\mathrm{c}}<d_{\mathrm{p}}$},
    \end{cases}
\end{align}
where $x_{\mathrm{p},q,i}$ and $x_{\mathrm{p},r,i}$ are the center positions of the two particles, $\delta_{\mathrm{c}}=|x_{\mathrm{p},q,i}-x_{\mathrm{p},r,i}|$ is the absolute distance between the two particles centers, and $\Delta r_{\mathrm{c}}=\Delta x$ is a threshold distance that is chosen equal to the fluid mesh spacing. The collision coefficient is chosen as 
\begin{align}
    \kappa_{\mathrm{c}} = \rho_{\mathrm{p}}V_{\mathrm{p}}|g|,
\end{align}
where $V_{\mathrm{p}}$ is the volume of each particle. Note that \citet{Willen2019} treat particle collisions differently than in the present case. However, since the mean particle volume fraction is with $\langle \epsilon_{\mathrm{p}}\rangle=0.087$ relatively small, we expect the present configuration to be dominated by fluid-particle interactions instead of particle-particle interactions. \\
In order to achieve a statistically stationary state of the particles settling in the fully periodic domain, a force must be applied to the fluid that counteracts the difference between the gravitational force and the buoyancy force acting on the particles, such that the total external force applied to the fluid-particle system is zero. Considering $N_{\mathrm{p}}$ particles each with a volume $V_{\mathrm{p}}$, the force field per unit volume in the vertical direction applied to the fluid is given as 
\begin{align}
    s_{\mathrm{f}}(\boldsymbol{x}) = \dfrac{(\rho_{\mathrm{p}}-\rho_{\mathrm{f}})N_{\mathrm{p}}V_{\mathrm{p}}|g|}{V_{\mathrm{f}}} \epsilon_{\mathrm{f}}(\boldsymbol{x}),
\end{align}
where $V_{\mathrm{f}}$ is the total volume of the fluid. The integral of the force field per unit volume, $s_{\mathrm{f}}$, over the entire domain equals the force applied to the particles
\begin{align}
    \int_{\Omega} s_{\mathrm{f}}(\boldsymbol{x}) \mathrm{d}V = (\rho_{\mathrm{p}}-\rho_{\mathrm{f}})N_{\mathrm{p}}V_{\mathrm{p}}|g|.
\end{align}
Therefore, the total momentum of the fluid-particle system is conserved, but a mean upward fluid velocity and a mean downward particle velocity develop. \\
\begin{table}[h]
    \centering
     \caption{List of the simulation parameters for the periodic settling of spheres.}
    \label{tab:parametersperiodicsedimentation}
    \begin{tabular}{ccccccc}
    \toprule
    $\mathrm{Re}_{\mathrm{p}}$ & $\mathrm{Ga}$ & $\rho_{\mathrm{p}}/\rho_{\mathrm{f}}$ & $L_{\mathrm{h}}/d_{\mathrm{p}}$ & $L_{\mathrm{v}}/d_{\mathrm{p}}$ & $N_{\mathrm{p}}$ & $\langle \epsilon_{\mathrm{p}} \rangle$ \\
    \midrule
    $110.8$ & $99.4$ & $5$ & $10$ & $30$ & $500$ & $0.087$\\
    \bottomrule
    \end{tabular}
\end{table}
The physical parameters of the present configuration are summarized in table \ref{tab:parametersperiodicsedimentation}. The particle Reynolds number is computed using the terminal velocity of a single settling particle, $v_{\mathrm{t}}$. Note that the simulation of the single settling particle of the present case is not performed in the present paper, but the terminal velocity is extracted from the simulation performed by \citet{Willen2019}. The Galilei number of the present configuration is given as 
\begin{align}
    \mathrm{Ga} = \dfrac{\rho_{\mathrm{f}}}{\mu_{\mathrm{f}}}\sqrt{\left( \dfrac{\rho_{\mathrm{p}}}{\rho_{\mathrm{f}}} -1 \right)d_{\mathrm{p}}^3|g|}.
\end{align}
Figure \ref{fig:periodicsettlingresults} shows the mean vertical particle velocity relative to the vertical superficial fluid velocity (average volume-filtered fluid velocity), $(\langle v_{\mathrm{v}}\rangle-\langle \epsilon_{\mathrm{f}}\Bar{u}_{\mathrm{v}} \rangle)$, and the vertical and horizontal particle velocity fluctuations of the \newIBM{} with spatial resolutions of $d_{\mathrm{p}}/\Delta x=8$ and $d_{\mathrm{p}}/\Delta x=12$ compared to the results reported in \citet{Willen2019}. The number of surface triangles per particle used in the simulations is $N_{\mathrm{s}}=320$. The particle velocity fluctuations are defined as deviations from the mean particle velocity, $v^{\prime}=v-\langle v\rangle$. Note that the horizontal values reported are the sum of the mean squared fluctuations of the two horizontal directions. The reference values for normalization are defined as in \citet{Willen2019}, and, are given as 
\begin{align}
    v_{\mathrm{ref}}&=(1-\langle \epsilon_{\mathrm{p}} \rangle)^{n-1}|v_{\mathrm{t}}|, \\
    t_{\mathrm{ref}}&=d_{\mathrm{p}}/v_{\mathrm{ref}},
\end{align}
where the exponent is $n=3.003$. \\
The mean relative vertical velocity between fluid and particles, as shown in figure \ref{fig:periodicsettlingresults}, increases rapidly after the initial condition, where the fluid and the particles are at rest. After the initial acceleration, the relative vertical velocity obtained with the two \newIBM{} resolutions reaches a relatively steady state with small fluctuations. The magnitude of the relative velocity predicted by the \newIBM{} is smaller than the one reported by \citet{Willen2019} for both resolutions. While the \newIBM{} with a resolution of $d_{\mathrm{p}}/\Delta x=8$ predicts an error of approximately 10\%, the \newIBM{} with a resolution of $d_{\mathrm{p}}/\Delta x=12$ predicts the reference relative velocity with good accuracy.  \\
The statistics of the correlations of the particle velocity fluctuations, as shown in figure \ref{fig:periodicsettlingresults}, are not fully converged, such that a significantly longer simulation time than $200t_{\mathrm{ref}}$ seems to be required to compute reliable mean correlations. However, the results for the present simulation time suggest that \newIBM{} predicts larger correlations in both directions than the correlations reported in \citet{Willen2019} but the \newIBM{} with the higher resolution of $d_{\mathrm{p}}/\Delta x=12$ predicts both correlation of the particle velocity fluctuations reasonably well.

%% file: Sections/conclusions.tex
\section{Conclusions}
\label{sec:conclusions}
In the present paper, the immersed boundary method (IBM) is revisited using the approach of volume-filtering the Navier-Stokes equations (NSE). Recent advances in volume-filtering enable reliable closure modeling of the volume-filtered NSE for small and moderate filter widths \cite{Hausmann2024a}. It is demonstrated in the present paper that the discretization of the particle momentum source arising in the volume-filtered NSE is the basis for a numerical methodology that is closely related to \oldIBM{} with continuous forcing. However, the volume-filtering framework leads to essential differences compared to the \oldIBM{} in a number of aspects. Most importantly: (\romannumeral 1) The volume-filtered fluid velocity differs from the particle velocity at the surface, which generally causes a volume-filtered flow passing through the particle surface. (\romannumeral 2) The governing equations for the fluid flow are not the NSE but the volume-filtered NSE that contain closure terms arising from the unresolved subfilter scales. A \oldIBM{} with continuous forcing does not constitute an accurate physical basis for the flow inside the particle and for spreading the force across multiple cells. The volume-filtering framework, however, inhibits a physical foundation for the spreading of the force and the flow field inside the particle, which is why we refer to the new framework as \newIBM{}. \\
In addition to the physical justification of the IBM framework, the \newIBM{} enables predictions of the forces acting on particles with good accuracy, even for spatial resolutions coarser than $d_{\mathrm{p}}/\Delta x=10$, at which the \oldIBM{} typically fails to predict reasonable results. \\
A number of validation cases are studied to assess the predictions of the newly proposed \newIBM{}. (\romannumeral 1) In the case of a uniform flow around a fixed isolated sphere, a spatial resolution of only $d_{\mathrm{p}}/\Delta x=6$ is required with the \newIBM{} to predict the drag force on the particle with less than 10\% deviation for the investigated particle Reynolds numbers in the range $10\leq \mathrm{Re}_{\mathrm{p}}\leq250$. (\romannumeral 2) In different arrangements of three closely packed spheres, the \newIBM{} requires a spatial resolution of $d_{\mathrm{p}}/\Delta x=8$ to predict less than 10\% error of the drag coefficient. (\romannumeral 3) In a configuration with 123 randomly packed particles with an average particle volume fractions of $\langle \epsilon_{\mathrm{p}}\rangle=0.2$, approximately 28\% of the predicted forces on the particles have an error that exceeds 10\% when using the \newIBM{} with a spatial resolution of $d_{\mathrm{p}}/\Delta x=8.75$. It can be concluded that a slightly higher spatial resolution is required with the \newIBM{} in flows with large particle volume fractions compared to the isolated sphere configuration to obtain similar accuracy of the predicted forces on the particles. (\romannumeral 4) The accurate computation of the contributions of the closure terms, i.e., the viscous closure and the subfilter stress closure in the volume-filtered NSE, is assessed by considering the settling of an isolated sphere. The terminal velocity of the particle predicted by the \newIBM{} with a spatial resolution of $d_{\mathrm{p}}/\Delta x=7.5$ deviates approximately 6\% from the terminal velocity of the experiments carried out by \citet{Mordant2000}. (\romannumeral 5) In the configuration of periodic settling of 500 spheres, a spatial resolution of $d_{\mathrm{p}}/\Delta x=12$ is shown to be sufficient with the \newIBM{} to reproduce the mean settling velocity and the particle velocity fluctuations of the periodic settling configuration from \citet{Willen2019} accurately. \\
The investigated validation configurations show that the \newIBM{} with a coarse spatial resolution compared to \oldIBM{} and, therefore, significantly smaller computational costs, is capable of approximating the fluid forces on particles with an accuracy that is sufficient for many applications. The presented methodology presents a step towards closing the gap between either fully resolving the fluid-particle interactions with high spatial resolutions and point-particle modeling, which is typically accurate only for coarse resolutions \cite{Evrard2021}.




%% file: Sections/acknowledgements.tex
\section*{Acknowledgements}
This work was funded by the Deutsche Forschungsgemeinschaft (DFG, German Research Foundation) - Project-ID 422037413 - TRR 287.

%% file: Sections/appendix.tex
\appendix


\section{Volume-filtered fluid velocity at the particle surface in the limit of zero filter width}
\label{ap:zerofilterwidthlimit}
The analytical expression for the fluid volume fraction of the particle with the index $q$ assuming a Gaussian filter with a standard deviation, $\sigma$, is given as \cite{Balachandar2022}
\begin{align}
    \epsilon_{\mathrm{p},q}(\boldsymbol{x}) = \dfrac{\mathrm{erf}(A)-\mathrm{erf}(B)}{2}+\dfrac{\exp(-A^2) - \exp(-B^2)}{\sqrt{\pi} (A+B)},
\end{align}
with 
\begin{align}
    A=\dfrac{2 |\boldsymbol{x}-\boldsymbol{x}_{\mathrm{p},q}|+d_{\mathrm{p},q}}{2\sqrt{2}\sigma}, \\
    B=\dfrac{2 |\boldsymbol{x}-\boldsymbol{x}_{\mathrm{p},q}|-d_{\mathrm{p},q}}{2\sqrt{2}\sigma},
\end{align}
and the particle diameter, $d_{\mathrm{p},q}$. The particle volume fraction at the surface of the particle converges to $\epsilon_{\mathrm{p},q}(\boldsymbol{X}_l)=0.5$ as $\sigma\rightarrow 0$ since
\begin{align}
    A&\rightarrow \infty,\\
    B&= 0.
\end{align}
Consequently, the fluid volume fraction at the particle converges to $\epsilon_{\mathrm{p},q}(\boldsymbol{X}_l)=0.5$. Assuming the no-slip boundary condition for the fluid velocity at the particle surface, $\boldsymbol{u}(\boldsymbol{X}_l)=\boldsymbol{u}_{\mathrm{Int},l}$, the volume-filtered fluid velocity at the particle surface with $\sigma \rightarrow 0$ is given as
\begin{align}
    \overline{I_{\mathrm{f}}u_i}\vert_{\sigma\rightarrow0} = \dfrac{1}{2}u_{\mathrm{Int},l,i}.
\end{align}

\section{Analytical expression for $\mathcal{L}$ for spherical particles}
\label{ap:analyticallengthscale}
In the case of a spherical particle, the integral of the Gaussian kernel, $g_{\mathrm{L}}$, over the surface of the particle can be computed analytically, such that the length scale $\mathcal{L}$ is given as
\begin{align}
    \dfrac{1}{\mathcal{L}(\boldsymbol{X}_l)} = \sum_q \int\displaylimits_{\partial\Omega_{\mathrm{p},q}}g_{\mathrm{L}}(|\boldsymbol{x}-\boldsymbol{y}|)\mathrm{d}A_y = \sum_q\dfrac{a_q \exp{\left(-\dfrac{(a_q+r_{l,q})^2}{2\sigma_{\mathrm{L}}^2}\right)}}{\sqrt{2 \pi \sigma_{\mathrm{L}}^2}r_{l,q}} \left( \exp{\left(\dfrac{2a_q r_{l,q}}{\sigma_{\mathrm{L}}^2}\right)} - 1\right),
\end{align}
where $a_q$ is the radius of the spherical particle with the index $q$ and $r_{q,l}=|\boldsymbol{X}_l - \boldsymbol{x}_{\mathrm{p},q}|$ is the Euclidean distance between the particle center, $\boldsymbol{x}_{\mathrm{p},q}$, and the considered Lagrangian surface marker, $\boldsymbol{X}_l$.